\documentclass[a4paper,fleqn,usenatbib]{mnras}
\usepackage{mathptmx}
\usepackage[T1]{fontenc}
\usepackage{ae,aecompl}
\usepackage{times}
\usepackage{amssymb}
\usepackage{amsmath}
\usepackage{upgreek}
\usepackage{color}
\usepackage{graphicx}
\usepackage{pdflscape}
\usepackage{multicol} 

\title[On the formation of the "L" type filament in G286.21+0.17]{ATOMS: ALMA Three-millimeter Observations of Massive Star-forming regions -- VI. On the formation of the "L" type filament in G286.21+0.17}

\author[J. W. Zhou, T. Liu, J.-Z. Li et al.]{
Jian-Wen Zhou,\thanks{E-mail: zjw@nao.cas.cn }$^{1,2}$
Tie Liu,\thanks{E-mail: liutie@shao.ac.cn }$^{3,4}$
Jin-Zeng Li,\thanks{E-mail: ljz@nao.cas.cn}$^{1}$
Hong-Li Liu,$^{5,6}$
Ke Wang,$^{7}$
Feng-Wei Xu,$^{8}$
Kee-Tae Kim,$^{9,10}$
\newauthor
Chang Won Lee,$^{9,10}$
Lokesh Dewangan,$^{11}$
Ken'ichi Tatematsu,$^{12}$
Shanghuo Li,$^{9}$
Xun-Chuan Liu,$^{8}$
\newauthor
Mengyao Tang,$^{14}$
Zhiyuan Ren,$^{1}$
Guo-Yin Zhang,$^{1}$
Chao Zhang,$^{5}$
Rong Liu,$^{1,2}$
Qiu-Yi Luo,$^{3,2}$
\newauthor
Isabelle Ristorcelli$^{13}$
\\
Affiliations are listed at the end of the paper}
\date{Accepted XXX. Received YYY; in original form ZZZ}

\pubyear{2021}

\begin{document}
\label{firstpage}
\pagerange{\pageref{firstpage}--\pageref{lastpage}}
\maketitle

\begin{abstract}
Filaments play an important role in star formation, but the formation process of filaments themselves is still unclear. The high-mass star forming clump G286.21+0.17 (G286 for short) that contains an "L" type filament was thought to undergo global collapse. Our high resolution ALMA band 3 observations resolve the gas kinematics of G286 and reveal two sub-clumps with very different velocities inside it. We find that the "blue profile" (an indicator of gas infall) of HCO$^+$ lines in single dish observations of G286 is actually caused by gas emission from the two sub-clumps rather than gas infall. We advise great caution in interpreting gas kinematics (e.g., infall) from line profiles toward distant massive clumps in single dish observations. Energetic outflows are identified in G286 but the outflows are not strong enough to drive expansion of the two sub-clumps. The two parts of the "L" type filament (“NW-SE” and “NE-SW” filaments) show prominent velocity gradients perpendicular to their major axes, indicating that they are likely formed due to large-scale compression flows. We argue that the large-scale compression flows could be induced by the expansion of nearby giant H{\sc ii} regions. The “NW-SE” and “NE-SW” filaments seem to be in collision, and a large amount of gas has been accumulated in the junction region where the most massive core G286c1 forms. 
\end{abstract}


%%%%%%%%%%%%%%%%%%%%%%%%%%%%%%%%%%%%%%%%%%%%%%%%%%

%%%%%%%%%%%%%%%%% BODY OF PAPER %%%%%%%%%%%%%%%%%%

\section{Introduction} \label{1}
High-mass stars that generally form in clusters are born in 
dense molecular clumps. However, the conversion of turbulent, self-gravitating, dense molecular gas clumps into high-mass stars as well as star clusters remains an open question in astrophysics \citep{Zinnecker2007,Motte2018,Jackson2019}. At present, there are several theoretical models for high-mass star formation.
The “competitive accretion” model \citep{Bonnell1997,Bonnell2001} explicitly relies on the funneling of gas into the center of the clump in order to provide fresh material for the central cores to accumulate enough mass to build up a high-mass star. In other words, this model predicts that the clump must be undergoing global collapse. The “global hierarchical collapse” model \citep{Vazquez-Semadeni2009,Vazquez-Semadeni2017} is an extension of the competitive accretion model when accretion through inflowing gas streams driven by gravity replaces the Bondi–Hoyle accretion \citep{Smith2009}. In this scenario, high-mass protostars would then be fed from the gas of their surrounding massive dense cores (MDCs)/clumps \citep{Motte2018}.
Although the “turbulent core accretion” model \citep{McKee2003,Krumholz2005} has no such explicit requirement, the formation of the parsec-scale dense clump itself almost certainly results from the gravitational collapse of a larger giant molecular cloud. Thus, for all of these theoretical models, large-scale (parsec-scale) infall motions due to gravitational collapse are expected \citep{Jackson2019}. Therefore, observations of gas infall motion in protoclusters are essential to test these models. Global collapse of protocluters or massive clumps has been suggested in some previous observations (e.g., \citealt{Barnes10,Liu2013,Peretto2013,He2015,Yuan2017}).

The existence of filaments in the interstellar medium has been known for a long time. In particular, the imaging surveys with the Herschel observatory have greatly promoted the studies of filamentary structures in molecular clouds \citep{Andre2010,Molinari10,Arzoumanian2011,Arzoumanian2019}.  Majority of the prestellar cores or protostars are found
to be embedded within such filamentary structures, indicating
that dense molecular filaments are the important sites of star formation \citep{Marsh2016, Konyves2015, Konyves2020}. Filaments will fragment into dense cores when the filaments reach a critical density or line-mass \citep{Vazquez-Semadeni1994, Krumholz2007, Zhang2020}. Magnetic fields also play an important role in stabilizing filaments \citep{Fiege2000, Hennebelle2013,Gomez2018}. Although the role of filaments in star formation is increasingly confirmed by observations,  the formation process of filaments themselves is still unclear \citep{Andre2014}. 

G286.21+0.17 (hereafter G286) is a massive protocluster site associated with the $\eta$ Car giant molecular cloud at a distance of $2.5\pm0.3\:$kpc, in the Carina spiral arm (e.g., \citealt{Barnes10}, hereafter B10; \citealt{Andersen2017}). Through observations of molecular lines, B10 determined several parameters about G286 region: the size $\sim0.9$ pc, the luminosity $\sim(2-3)\times10^{4}\:L_\odot$, and the mass $\sim20 000\:M_\odot$. From the line profiles of the J=1-0 and J=4-3 transitions of HCO$^{+}$ and H${^{13}}{\rm C}$O$^{+}$ line emission, B10 argued that G286 is undergoing global collapse and they found one of the largest such infall rates yet measured in protoclusters, $\sim3\times10^{-2}\:M_\odot\:{\rm{yr}}^{-1}$. From {\it Herschel} column density map, \cite{Ohlendorf2013} determined the mass of the cloud is $\sim2 105\:M_\odot$ by integrating over a 2.23 pc $\times$ 2.23 pc box around the clump. The densest part of G286 region, or the dense clump associated with the infrared source IRAS 10365–5803, has a mass of only $\sim470\:M_\odot$ \citep{Faundez2004}. This work is focused on the dense clump.

\citet{Andersen2017} suggests the cluster are very young with a rich population of pre-main-sequence stars with a disk fraction ranging from 27$\%$ to 44$\%$ for a magnitude and extinction limited sample. \citet{Cheng2020ApJ} presents a near-infrared (NIR) variability analysis toward the G286  protocluster, suggesting it as being an extreme case of an outburst event that is still ongoing. These properties make G286 a particularly interesting site of high-star formation. \citet{Cheng2018} present the 1.3 mm ALMA continuum observation of G286 and an analysis of the core mass function (CMF) in this region. Subsequently, \citet{Cheng20} (hereafter C20) present a follow-up study of multiple spectral lines from the same ALMA observations to investigate the gas kinematics and dynamics of G286 from clump to core scales. In this paper, we use new ALMA data from the ATOMS survey (see Sec.~\ref{alma}) to revisit the G286 clump, focusing on testing the global infall  proposed by B10. We also discuss the outflows, filaments and star formation history in this region. 

\section{Observations} 
\label{sec:style}

\subsection{Single dish data}

The single-pointing observations with the Atacama Pathfinder Experiment (\citealt{Gusten2006}, APEX) in CO (4-3) and C$^{17}$O (3-2) lines were conducted from August 8th to 10th, 2015. All observations were performed using the position
switching mode, the pointing position is RA=10:38:32.00, Dec=-58:19:15.20. The reference position was carefully selected in a circular region with a diameter of 3 arcdeg centered at the “on” source position, which has no or extremely weak CO emission based on previous CO surveys of the Milky Way \citep{Dame2001}. The antenna system temperatures range from 236 to 255 K, and from 865 to 1145 K respectively, during C$^{17}$O (3-2) and CO (4-3) observations. The frontend used was the MPIFR dual channel FLASH receiver (460 and 810 GHz) and APEX2 receiver (345 GHz) \citep{Heyminck2006}, along with the APEX digital Fast Fourier Transform Spectrometer backends (\citealt{Klein2006}, FFTS). The lines were covered with 2048 channels within a bandwidth of 1 GHz. We have smoothed the spectra to 0.27 and 0.4 km s$^{-1}$ resolution for CO (4-3) and C$^{17}$O (3-2) lines. The mean rms noise levels are 0.19 and 0.14 K for CO (4–3) and C$^{17}$O (3–2) lines, respectively. The beam sizes are 13$\arcsec$ and 18$\arcsec$ for CO (4–3) and C$^{17}$O (3–2) lines, respectively. More details of the APEX observations were summarised in \cite{Yue2021}.

The single pointing observations of G286 with the Atacama Submillimeter Telescope Experiment (ASTE) 10 m telescope were conducted between 2014 July 1st and 4th \citep{Liu2016}. All observations were performed using the position
switching mode, the pointing position is RA=10:38:32.02, Dec=-58:19:15.10. The two-side band single-polarization heterodyne receiver DASH345/CATS345, operating at frequencies of 324–372 GHz was used to observe HCN (4-3) and CS (7-6) lines simultaneously. We used the MAC spectrometer with a spectral resolution of 0.5 MHz or 0.42 km s$^{-1}$. The beam size at 354 GHz is $\sim22\arcsec$. The main beam efficiency was $\sim0.6$. The system temperature varied from 400 to 500 K during observations. The rms level per channel is $\sim$0.1-0.2 K in antenna temperature. More details of the ASTE observations were summarised in \cite{Liu2016}.

\subsection{ALMA observations}\label{alma}
We use ALMA data from the ATOMS survey (Project ID: 2019.1.00685.S; PI: Tie Liu). The ACA 7m observations of G286 (or I10365-5803) were conducted on the 19th October, 2019. The 12-m array observations of G286 were conducted on the 2nd November, 2019. The typical 12-m array time on source is $\sim$ 3 minutes, while typical ACA 7m observing time per source is $\sim$ 8 minutes. The angular resolution and maximum recovered angular scale for the ACA 7m observations are $\sim13\arcsec$ and $\sim73\arcsec$, respectively. The angular resolution and maximum recovering angular scale for the 12-m array observations are $\sim1.7\arcsec$ and $\sim19\arcsec$, respectively. The achieved angular resolutions of the 12-m array observations for molecular lines are usually better than 2$\arcsec$ (or $\sim$0.024 pc at 2.5 kpc distance), enabling us to resolve dense cores ($\sim0.1$ pc in size) in G286. More details of the ALMA observations were summarised in \cite{Liu2020,Liu2020b}.

Calibration and imaging were carried out using the CASA software package version 5.6 \citep{McMullin2007}. The ACA 7m data and ALMA 12m array data were calibrated separately. Then the visibility data from the ACA 7m and ALMA 12m array configurations were combined and later imaged in CASA. For each source, each spectral window (spw), a line-free frequency range is automatically determined using the ALMA pipeline (see ALMA technical handbook). This frequency range is used to (a) subtract continuum from line emission in the visibility domain, and (b) make continuum images. Continuum images are made from multi-frequency synthesis of data in this line-free frequency ranges in the two 1.875 GHz wide spectral windows, spw 7 and 8, centered on $\sim99.4$ GHz (or 3 mm). Visibility data from the ALMA 12m and ACA 7m arrays is jointly cleaned using task tclean in CASA 5.6. We used natural weighting and a multiscale deconvolver, for an optimized sensitivity and image quality. All images are primary-beam corrected. The continuum image reaches a 1 $\sigma$ rms noise of 0.21 mJy in a synthesized beam of $2.12\arcsec \times 2.29\arcsec$ (PA= -56.74$^{\circ}$). For molecular lines, we focus on HCO$^{+}$ (1-0), H$^{13}$CO$^{+}$ (1-0), HC$_{3}$N (11-10), SiO (2-1), SO (3(2)-2(1)), CCH (1-0) and CS (2-1), which are summarized in Table \ref{line} for the 7m+12m combined data.

\begin{table}
	\centering
	\caption{Main targeted molecular lines in ALMA 7m+12m combined data}.
	\label{line}
	\begin{tabular}{cccc} % four columns, alignment for each
		\hline
Molecule & rms levels &spectral resolution &beam sizes\\
 &(Jy~beam$^{-1}$) &(km~s$^{-1}$) &\\
 \hline
HCO$^{+}$ (1-0) &0.012 &0.103 &$2.36\arcsec \times 2.53\arcsec$\\
H$^{13}$CO$^{+}$ (1-0) &0.008 &0.211 &$2.40\arcsec \times 2.61\arcsec$\\
HC$_{3}$N (11-10) &0.004 &1.463 &$2.09\arcsec \times 2.26\arcsec$\\
SiO (2-1) &0.008 &0.211 &$2.39\arcsec \times 2.63\arcsec$\\
SO (3(2)-2(1)) &0.004 &1.474 &$2.14\arcsec \times 2.33\arcsec$\\
CCH (1-0) &0.008 &0.210 &$2.39\arcsec \times 2.60\arcsec$\\
CS (2-1) &0.004 &1.494 &$2.14\arcsec \times 2.33\arcsec$\\
\hline
\multicolumn{4}{l}{}\\
	\end{tabular}
\end{table}

\subsection{Archived continuum emission data }
To study the physical characteristics of G286 clump and the interaction between the clump and its surrounding environment, auxiliary data of images were complemented by the GLIMPSE \citep{Benjamin2003},
MIPSGAL  \citep{Carey2009} and Herschel Infrared GALactic plane survey \citep{Molinari10,Molinari16}.
The images of the Spitzer Infrared Array Camera (IRAC) at 3.6, 4.5, 5.8, and 8.0 $\mu$m, together with
the Multiband Imaging Photometer for \emph{Spitzer} (MIPS) at 24 $\mu$m were retrieved from the
\emph{Spitzer} Archive. The angular resolutions of the images in the IRAC bands are $< 2\arcsec$ and it is
$\sim 6\arcsec$.0 in the MIPS 24 $\mu$m band.  The data of \emph{Herschel} Archive were obtained with the Photodetector Array Camera and Spectrometer (PACS, 70 and 160 $\mu$m) \citep{Poglitsch2010}, the measured angular resolutions of the data at these bands are 10$\farcs$7 and 13$\farcs$9 \citep{Traficante2011}. All of the Hi-GAL
data have been processed by PPMAP procedure, by which high resolution column density and dust temperature maps could be obtained. The resolution of column density and dust temperature maps is $\sim12\arcsec$, and the resulting data is available online\footnote{http://www.astro.cardiff.ac.uk/research/ViaLactea/} \citep{Marsh2017}. We also used ATLASGAL+Planck 870 $\mu$m data (ATLASGAL combined with Planck data), which are sensitive to a wide range of spatial scales at a resolution of $\sim21\arcsec$  \citep{Csengeri16}.

\section{Results} 
%{\color{red}
%\subsection{3mm continuum and infrared emission}
%In this section, display the 3 mm continuum images overlaid on %infrared images. Also plot the cores discovered by Cheng+2020 as %crosses.}
\subsection{Continuum emission}\label{3.1}

\begin{figure*}
\centering
\includegraphics[scale=0.5]{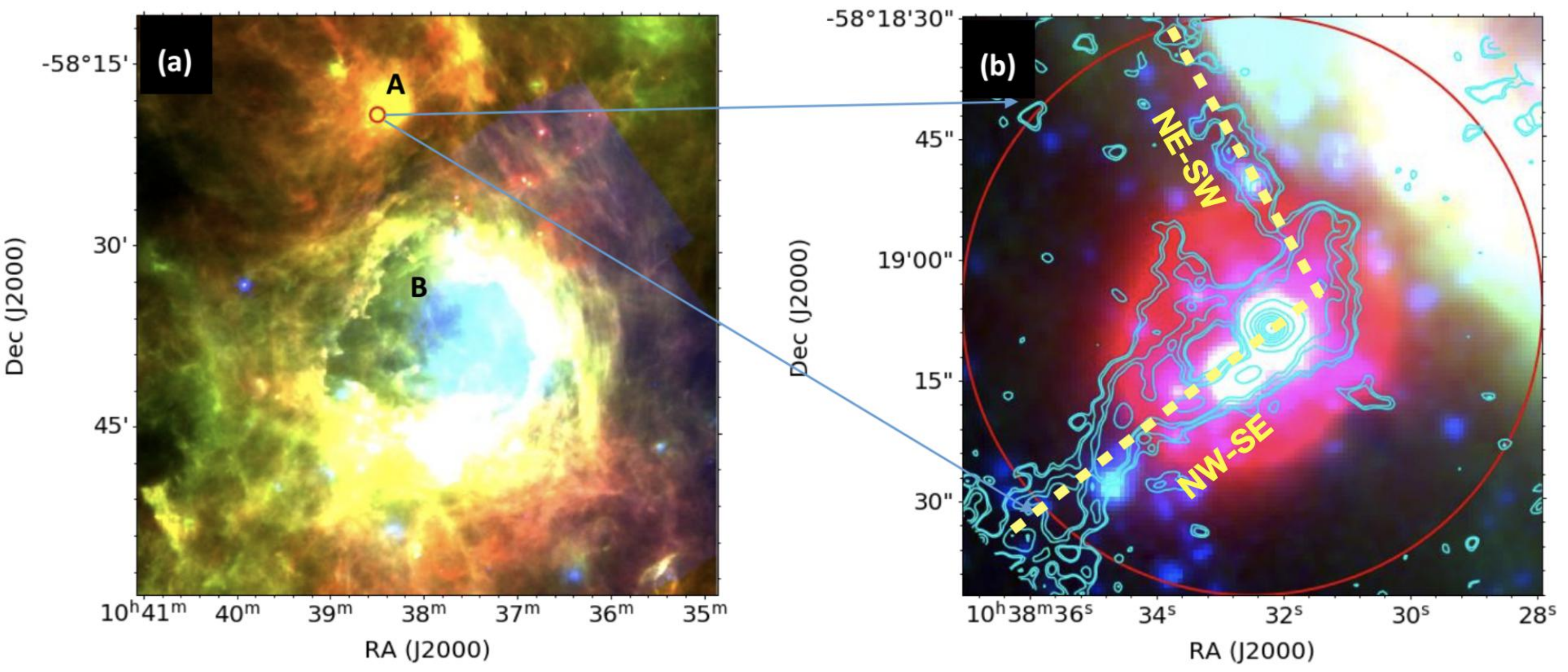}
\caption{(a) Three-color image of G286 and its surrounding environment constructed by combining Herschel PACS 160 (red), 70 (green) $\mu$m and Spitzer 24 $\mu$m (blue). Central coordinates of this figure is (Ra, Dec)=($159.50^\circ$, $-58.58^\circ$). Area of this region is $0.8^\circ$ $\times$ $0.8^\circ$. Red circle is the FOV of the ALMA 12m array observations; (b) Three-color image of the region inside red circle constructed by combining Spitzer IRAC 4.5 (blue), 5.8 (green) $\mu$m and Herschel PACS 70 $\mu$m (red). Cyan contours show the 3 mm continuum image combined ALMA 12 and 7 m array data. Contour levels are 1$\sigma$ $\times$ (1.5, 2, 4, 8, 16, 32, 64) with $\sigma$= 0.21 mJy/beam. The two filaments "NE-SW" and "NW-SE" are marked with yellow dashed lines.}
\label{red}
\end{figure*}

\begin{figure*}
\centering
\includegraphics[scale=0.35]{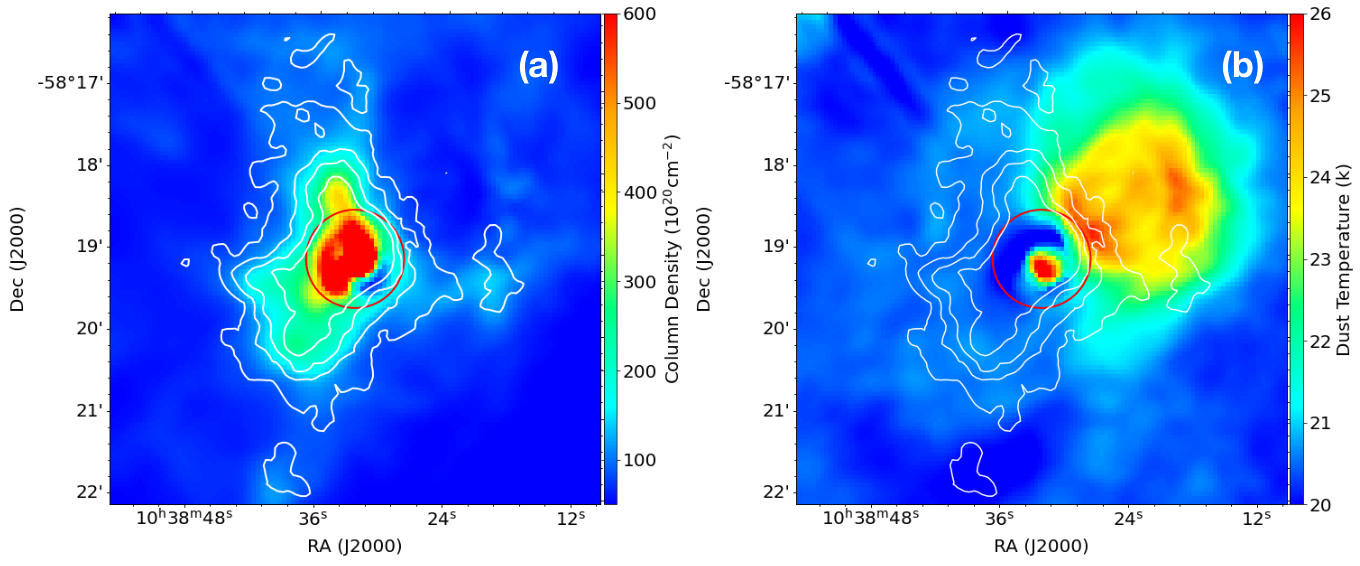}
\caption{
Column density map and temperature distribution map derived with the PPMAP procedure, the size of region is $6\arcmin \times 6\arcmin$, red circle is the FOV of the ALMA 12m array observations, white contour is the data of ATLASGAL+Planck 870$\mu$m. Contour levels are 1$\sigma$ $\times$ (3, 4.5, 6, 8) with $\sigma$= 0.20 Jy/beam.}
\label{sed}
\end{figure*}

Fig.~\ref{red}(a) presents a large-scale composite color map of the natal molecular cloud of G286 clump. Fig.~\ref{red}(b) is the blow-up image of G286 clump. The red circles in both panels (a) and (b) show the FOV of the ALMA 12-m array observations. Our ALMA observations cover the central region of the G286 clump. Fig.~\ref{sed} shows the column density and dust temperature maps of G286 clump derived with the PPMAP procedure. The average dust temperature over the central region of G286 clump (i.e, inside the red circle with a radius of R$\sim$0.44 pc, which is roughly the FWHM radius of the ALMA 12m-array FOV) is $\sim21.4$ K. The average column density N within the circle is about $5.66\times10^{22}$ cm$^{-2}$. Assuming the spherical geometry of clump, the number density is n= N/2R, then the mass of the central region in G286 clump is:
\begin{equation}
M =\frac{4}{3}\pi R^{3}\mu m_{H} n
\end{equation}
,where $\mu = 2.37$ is the mean molecular weight per ``free particle'' (H$_2$ and He, the number of metal particles is negligible), m$_{\rm H}$ and n are atomic hydrogen mass and molecular hydrogen number density. We obtain M $\sim$ 1021\:M$_{\sun}$.

A clear temperature gradient appears in the northwest-southeast direction of the clump in Fig.~\ref{sed}(b), indicating that the G286 clump is externally heated by the H{\sc ii} region A. The H{\sc ii} region A is visible as a Br-$\gamma$ emission nebula and is exactly coincident with a centimetre-continuum point source (see B10 for details). Protostars have formed inside the G286 clump, which are clearly seen as point sources in Spitzer images. Based on the previous studies, G286 clump is on the edge of H{\sc ii} region A (see B10). Both G286 and HII region A are found to be near H{\sc ii} region B (Gum31), which is at the same distance (d$\sim$2.5 kpc) and is excited by a young stellar cluster NGC3324 \citep{Cappa2008, Barnes10, Ohlendorf2013, Andersen2017}. H{\sc ii} region B is strongly interacting with the surrounding interstellar medium, and it may likely also affect the evolution of G286 and H{\sc ii} region A. In Sec.~\ref{4.3}, we will discuss in detail the effect of H{\sc ii} region A and B on star formation history in the G286 clump.

\begin{figure}
\centering
\includegraphics[scale=0.30]{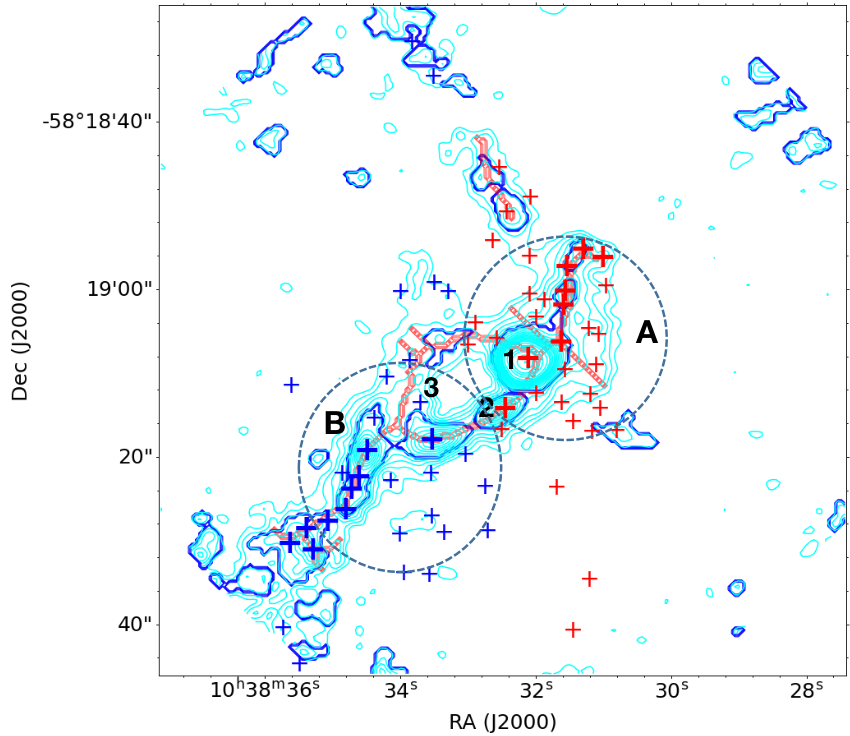}
\caption{3 mm continuum emission from 12m+7m data is shown as cyan contours with contour levels 1$\sigma$ $\times$ (1.5, 2, 2.8, 3.6, 4.4 to 12 by 0.4 steps, 12 to 15.2 by 0.8 steps, 18, 24, 30, 36, 48, 60) with $\sigma$=0.21 mJy/beam, and blue contours are the clumps found by Dendrogram algorithm. Red line is the skeleton of filament recognized by FILFINDER algorithm from 3 mm continuum emission. ''+'' represents dense cores identified by C20, and bold ”+” indicates the cores on the longest filament. Blue dashed circles mark two sub-clumps A and B identified in Fig.~\ref{cluster}. Red  ”+” show cores of sub-clump A, blue ”+” for sub-clump B.
The black number represents the serial number of the core corresponding to "+" (G286c1, G286c2 and G286c3).}
\label{fildc}
\end{figure}

As shown in the Fig.~\ref{red}(b), the main structure of the G286 clump in ALMA 3 mm continuum emission appears as “L” shape, which is labeled as “NE-SW” and “NW-SE” filaments in Fig.1 of C20. Fig.~\ref{fildc} shows the 3 mm continuum emission. The blue contours show compact cores extracted from the 3 mm continuum emission with the Dendrogram algorithm\footnote{https://dendrograms.readthedocs.io/en/stable/},  and the cores are generally distributed along the “L” structure. Red lines show the skeletons of sub-filaments identified with the FILFINDER algorithm \citep{Koch2015}. The length of longest sub-filament is $\sim0.77$ pc. The most massive cores, which are marked by bold "+" in Fig.~\ref{fildc}, are found to be located along the longest filament. We fit the radial profiles of longest sub-filament with the Radfil algorithm \citep{Zucker2018} and obtain the width (deconvolved FWHM) of longest filament, W$\sim$0.075 pc (the detailed description of the procedures is given in Appendix.~\ref{A}). 

%Based on the ALMA 3mm continuum emission flux density, we can roughly estimate the filament mass. The total flux and total mass of the whole G286 clump are $\sim0.196$ Jy and $\sim400$ M$_\odot$, respectively. The total flux of filaments is $\sim0.102$ Jy. Assuming that the dust temperature is constant within the clump, we can estimate the total mass of filaments from the flux ratio, $\sim208$ M$_\odot$, and thus the linear mass (M/l)$_{obs}$ $\sim270$ M$_\odot$ pc$^{-1}$. This value is comparable to the result of C20 ($\sim250$ M$_\odot$ pc$^{-1}$ for strip1).%

\begin{figure*}
\centering
\includegraphics[scale=0.35]{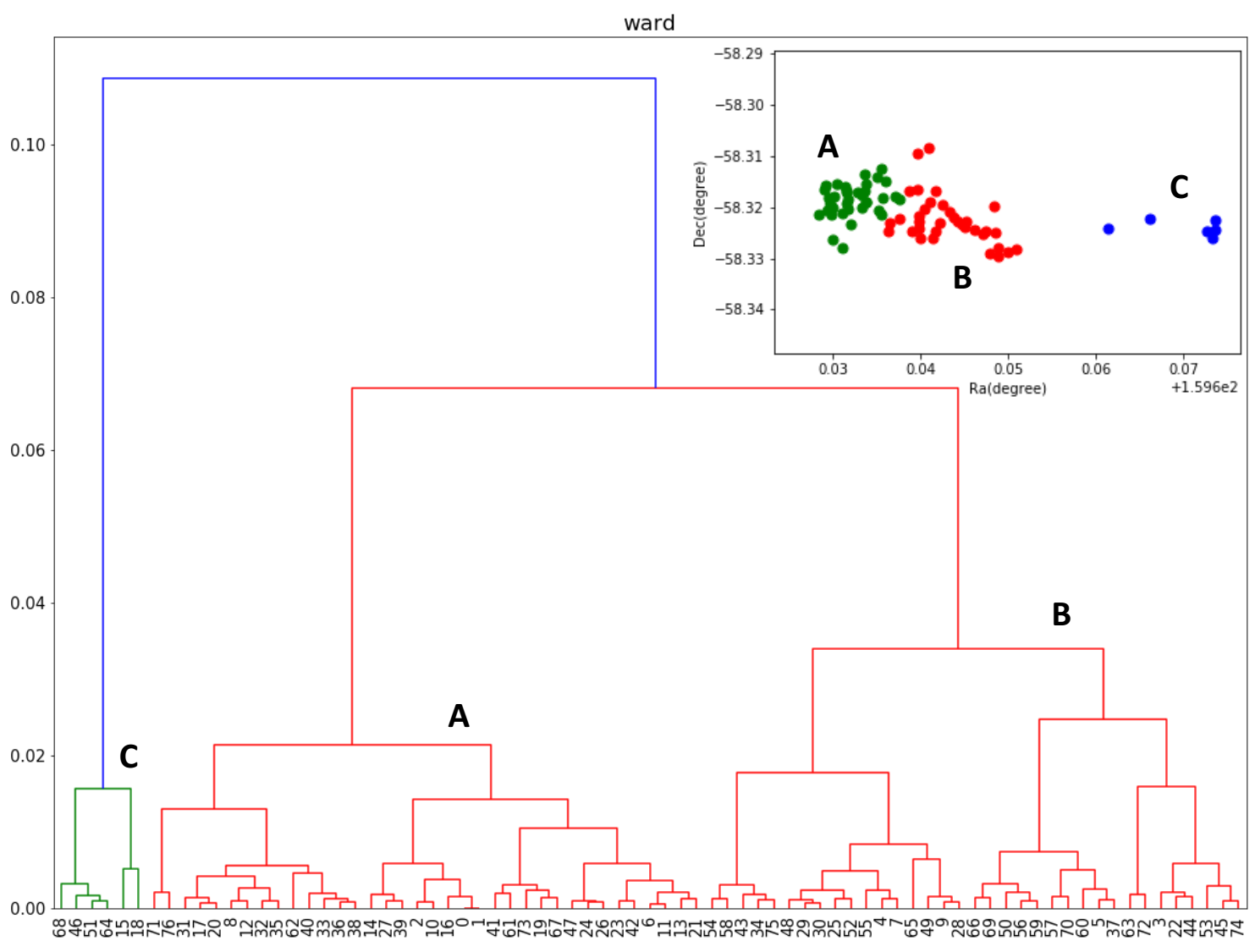}
\caption{Three clusters of 76 dense cores in G286 clump by applying a nearest neighbor separation (NNS) analysis.}
\label{cluster}
\end{figure*}

The 76 dense cores identified by 1.3mm continuum in C20 are marked by "+" in Fig.~\ref{fildc}. The 1.3mm continuum emission which has higher sensitivity reveals more fainter cores than the 3mm continuum emission. Fig.~\ref{cluster} shows a dendrogram tree obtained from the nearest neighbor separation (NNS) analysis to 76 dense cores (with the scipy function cluster.hierarchy.linkage and the “ward” method). The dense cores can be broadly divided into three clusters (A, B, C). Dense cores in cluster C (blue dots) are located beyond the FOV of our ALMA observations, and thus they are excluded in this study. The remaining clusters are named as sub-clump A and sub-clump B. In C20, they divided the cores into two groups "blue" and "red" based on their velocity differences. Interestingly, we find that sub-clumps A and B identified in the NNS analysis match the "red" group and "blue" groups by C20 well, respectively, suggesting that they are physically separated. We roughly mark the locations of two sub-clumps in Fig.~\ref{fildc} with blue and red dashed circles.

\subsection{Molecular line emission from single dish and ACA 7m observation}\label{3.2}
\begin{figure}
\centering
\includegraphics[scale=0.23]{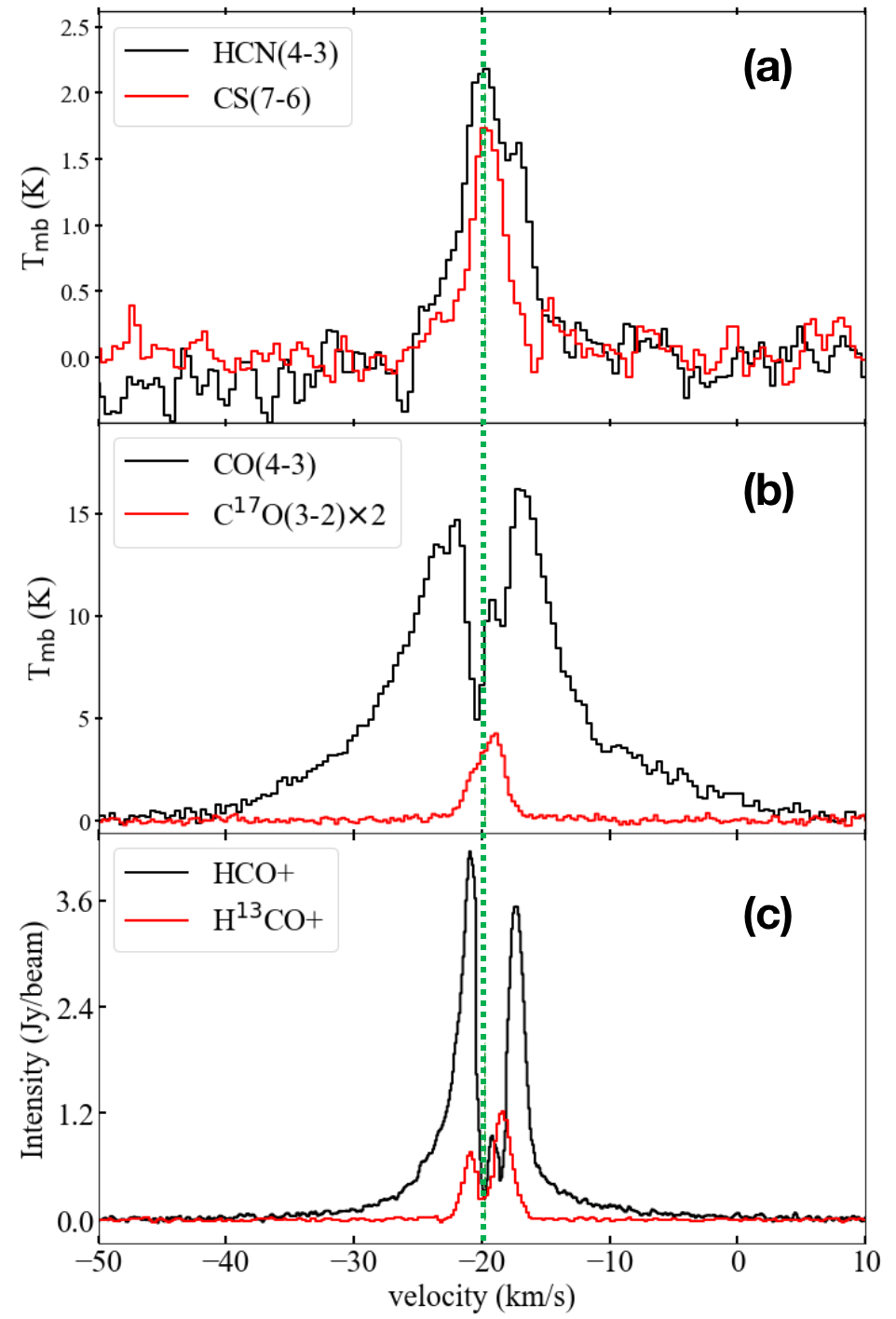}
\caption{(a) Averaged spectrum of HCN(4-3) (black) and CS(7-6) (red) for source G286 by ASTE; (b) Averaged spectrum for source G286 by APEX, black and green curves denote the CO(4-3) and C${^{17}}{\rm O}$ (3-2) lines; (c) HCO$^{+}$ (1-0) and H${^{13}}{\rm C}$O$^{+}$ (1-0) spectrum in G286 with ACA 7m array data averaged over a region with diameter $36\arcsec$ (the beam size of B10). Vertical dashed green line indicates the systemic velocity at V$_{LSR}$ = -19.8 km s$^{-1}$.}
\label{single}
\end{figure}

In B10, they found that the J=1-0 and J=4-3 of lines HCO$^+$ show double peak profiles with the blueshifted peak stronger than the redshifted one. Such "blue profile" was argued as evidence of large-scale gravitational infall in the dense gas.
Fig.~\ref{single}(a) shows the observation with ASTE in G286 region, the double-peak profile of HCN (4-3) is barely seen, but the line profile of CS (7-6) looks like a single-peak.
This is due to the poor spectral resolution ($\sim0.8$ km s$^{-1}$) in ASTE observations. The line profiles in ASTE observations are not well resolved. Moreover, we also observed G286 region with the APEX telescope (Fig.~\ref{single}(b)). The optically thick CO J=4-3 line shows a red profile rather than a blue profile, it also shows high velocity wing emission, suggesting the existence of energetic outflows in the G286 clump. The optically thin C${^{17}}{\rm O}$ J=3-2 line is single peaked but skewed to the redshifted side. 

The difference of the line profiles in ASTE and APEX observations could be related to the following two aspects: (1) Due to the poor spatial resolutions, single dish observations usually can not resolve the gas kinematics within massive clumps. Various line profiles could be caused by a mixture of gas kinematics (outflows, infall, rotation et al.) within a large beam. (2) Molecular lines with different critical densities or excitation conditions may trace various layers of the clump, resulting in different line profiles. 

Therefore, we advise great caution in interpreting gas kinematics (e.g., infall) from line profiles toward distant massive clumps in single dish observations.
%\subsection{Molecular line emission from ACA observations}\label{3.3}
Fig.~\ref{single}(c) presents the spectra of HCO$^{+}$ J=1-0 and H${^{13}}{\rm C}$O$^{+}$ J=1-0 from ACA 7m observations. The FOV of ACA 7m array is about two times larger than the ALMA 12-m array, and thus ACA 7m array data are useful to study gas kinematics at a larger scale. The spectra were averaged over a region with a size of $36\arcsec$, which is comparable to the Mopra beam size in B10. Both HCO$^{+}$ J=1-0 and H${^{13}}{\rm C}$O$^{+}$ J=1-0 show double peak profiles. The HCO$^{+}$ J=1-0 line profile in ACA 7m observations is blue profile, similar to the line profile in Mopra observations of B10. In B10, they argued that the optically thin line H${^{13}}{\rm C}$O$^{+}$ is single peaked. However, in ACA 7m observations, the H${^{13}}{\rm C}$O$^{+}$ J=1-0 line is double-peak profile. The moment maps of molecular line emission in ACA 7m array data are shown in Appendix.~\ref{B}.

\subsection{Molecular line emission from combined ALMA 12-m array and ACA data}\label{3.4}

\begin{figure}
\centering
\includegraphics[scale=0.18]{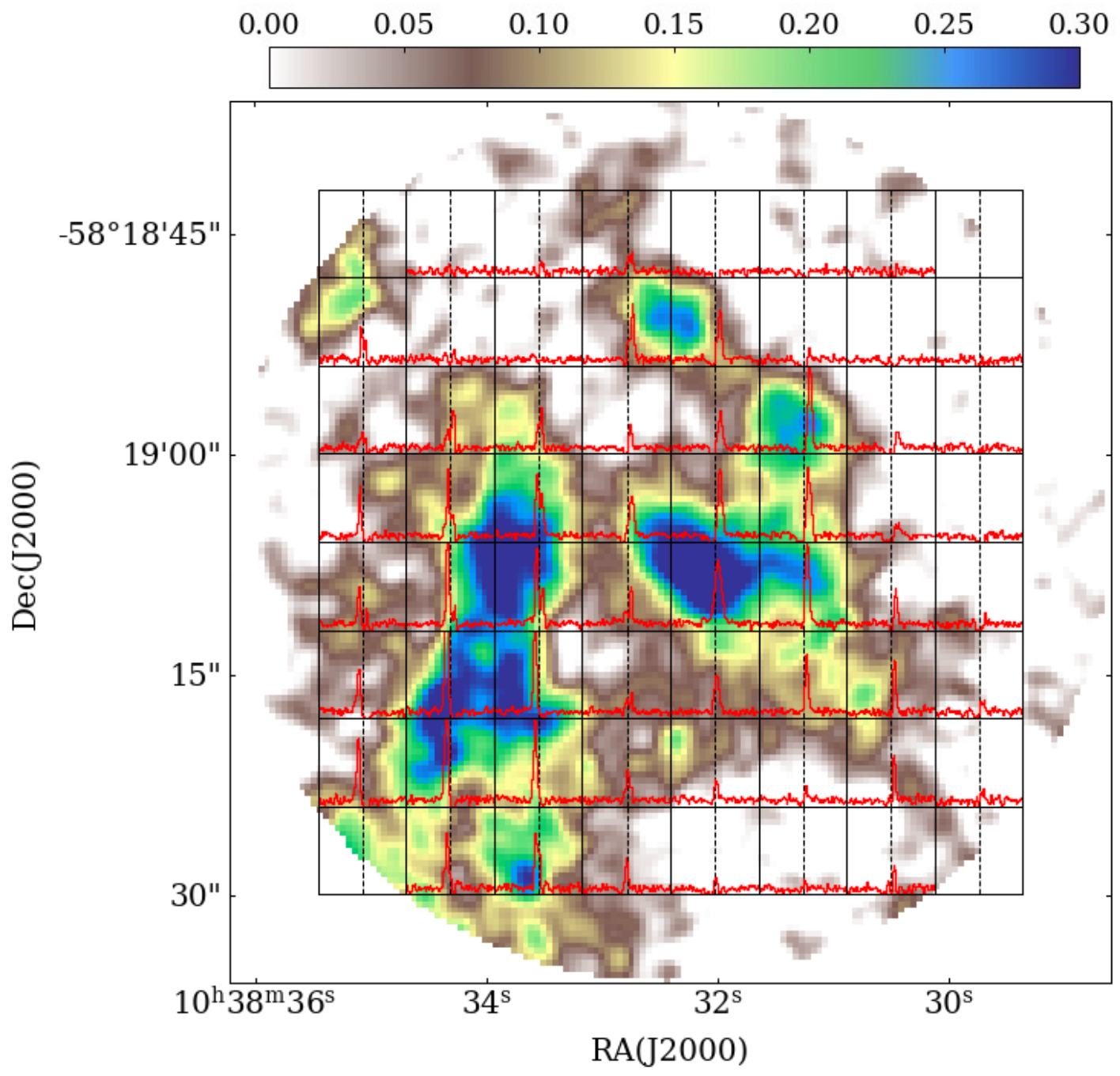}
\caption{The grid map of H${^{13}}{\rm C}$O$^{+}$ (1-0), background map is the moment 0 map of H${^{13}}{\rm C}$O$^{+}$, the side-length of each square lattice is 15 pixels. The line profile is the averaged one of the data in each square lattice, drawn in the velocity range of (-35 km s$^{-1}$, -5 km s$^{-1}$). Black dashed lines in the middle mark the systemic velocity. The color bar on the top indicates the flux scale in Jy beam$^{-1}$ km/s for moment 0 map.}
\label{h13cogrid}
\end{figure}

\begin{figure*}
\centering
\includegraphics[scale=2.1]{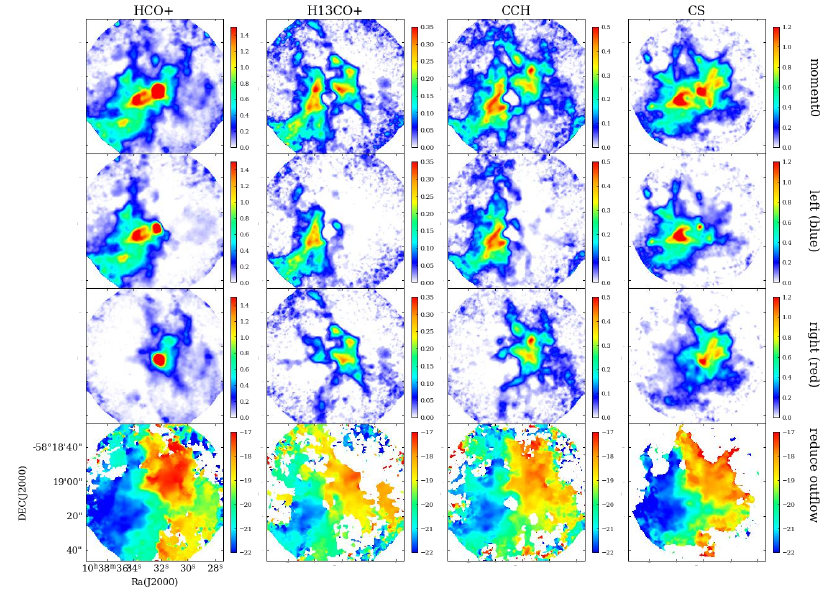}
\caption{Rows from left to right show the moment maps of HCO$^{+}$ (1-0), H$^{13}$CO$^{+}$ (1-0), CCH (1-0), and CS (2-1) with 7m+12m data. From top to bottom, the color scales show 7m+12m moment 0 maps of the whole [velocity range is (-60 km s$^{-1}$, 20 km s$^{-1}$)], 7m+12m moment 0 of the left [blueshift, velocity range is (-60 km s$^{-1}$, -19.8 km s$^{-1}$)], 7m+12m moment 0 of the right [redshift, velocity range is (-19.8 km s$^{-1}$, 20 km s$^{-1}$)], 7m+12m moment 1 maps of the whole reduced outflow. For eliminating outflow, the velocity range is limited as (-23 km s$^{-1}$, -16.5 km s$^{-1}$). The color bar on the right indicates the flux scale in Jy beam$^{-1}$ km/s for moment 0 maps and velocity scale in km s$^{-1}$ for moment 1 maps.}
\label{moment127}
\end{figure*}

To further study the gas kinematics in the G286 clump, we use the ALMA 12m+7m combined data for HCO$^{+}$ (1-0), H$^{13}$CO$^{+}$ (1-0), CCH (1-0), CS (2-1) and HC$_{3}$N (11-10). Fig.~\ref{h13cogrid} shows the grid map of H${^{13}}{\rm C}$O$^{+}$ (1-0). From left to right of the map, the emission peaks of  H${^{13}}{\rm C}$O$^{+}$ (1-0) change from redshift to blueshift with respect to the systemic velocity. The averaged spectra and grid maps of other lines are shown in Appendix.~\ref{B}, which having similar patterns to that of H${^{13}}{\rm C}$O$^{+}$ (1-0).

Figure \ref{moment127} presents the moment maps of HCO$^{+}$ (1-0), H$^{13}$CO$^{+}$ (1-0), CCH (1-0), and CS (2-1) in 7m+12m data. Two spatially separated sub-clumps are clearly seen from these moment maps. To reduce outflow contamination, we make moment 1 maps in the narrow velocity range of -23 km s$^{-1}$ to -16.5 km s$^{-1}$. The moment 1 maps shown in the last row of Figure \ref{moment127} also clearly reveal the large-scale velocity gradients along the north-west to south-east direction and the boundary of the two sub-clumps. These high resolution moment maps again confirm that the double peak profiles of molecular lines in single dish observations indeed come from the mixture of gas emission in two sub-clumps.

\begin{figure}
\centering
\includegraphics[scale=0.9]{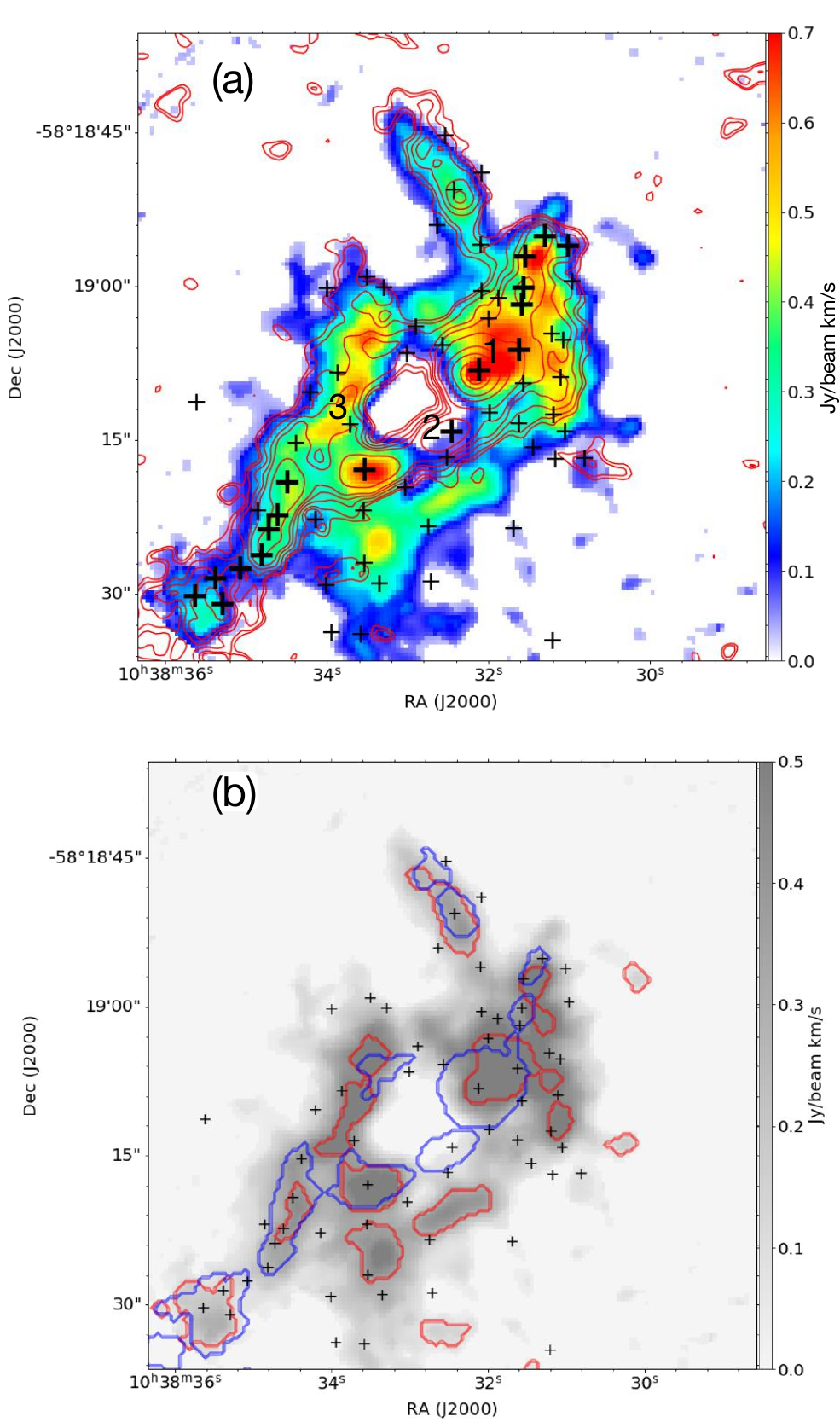}
\caption{(a) 7m+12m moment0 map of HC$_{3}$N (11-10) in color, red contours show the 3 mm continuum image for ALMA 12+7m array data. Contour levels are 1$\sigma$ $\times$ (1.5, 2, 3, 4, 5, 7.5, 15, 25, 45, 60) with $\sigma$= 0.21 mJy/beam; (b) Blue contours are the clumps found by dendrogram algorithm with 3mm continuum. Red contours are the clumps found by dendrogram algorithm with the moment0 map of HC$_{3}$N. Cross marks "+" represent dense cores identified by C20.}
\label{hc3n}
\end{figure}

HC$_{3}$N(11-10) is a good tracer of dense molecular gas \citep{Liu2020}. As shown in Fig.~\ref{hc3n}a, the morphology of HC$_{3}$N(11-10) emission matches with the 3 mm continuum emission very well. In addition, the clumps and/or cores found by dendrogram algorithm in 3mm continuum and in the moment 0 map of HC$_{3}$N match very well (see Fig.~\ref{hc3n}b). Hence, we suggest HC$_{3}$N can be a good tracer of filaments and dense cores. The two brightest cores (marked as "1" and "2" in Fig.~\ref{hc3n}a or Fig.~\ref{fildc}) are associated with two young stellar objects, which are clearly seen in Spitzer IRAC bands (Fig.~\ref{red}). Since the intensity sensitivity of HC$_{3}$N is higher than 3 mm continuum, the former reveals more structures than the latter. The “+” show the dense cores identified by C20 in 1.3 mm continuum. It can be seen that the dense cores traced by 1.3 mm emission do not always coincide well with HC$_{3}$N gas cores, which could be caused either by some chemical effect or different sensitivity between line and continuum data. In the northern part of G286, the HC$_{3}$N emission is embraced by the Spitzer 8 $\mu$m emission, which traces the PDR of the H{\sc ii} region to the north-west. It indicates that the H{\sc ii} region may greatly affect the dense gas distribution in the northern part of G286. Moreover, there is a cavity in the middle part of the moment 0 map of HC$_{3}$N, which is like the cavity blown out by outflows (see Sec.~\ref{3.5}).

\begin{figure}
\centering
\includegraphics[scale=0.4]{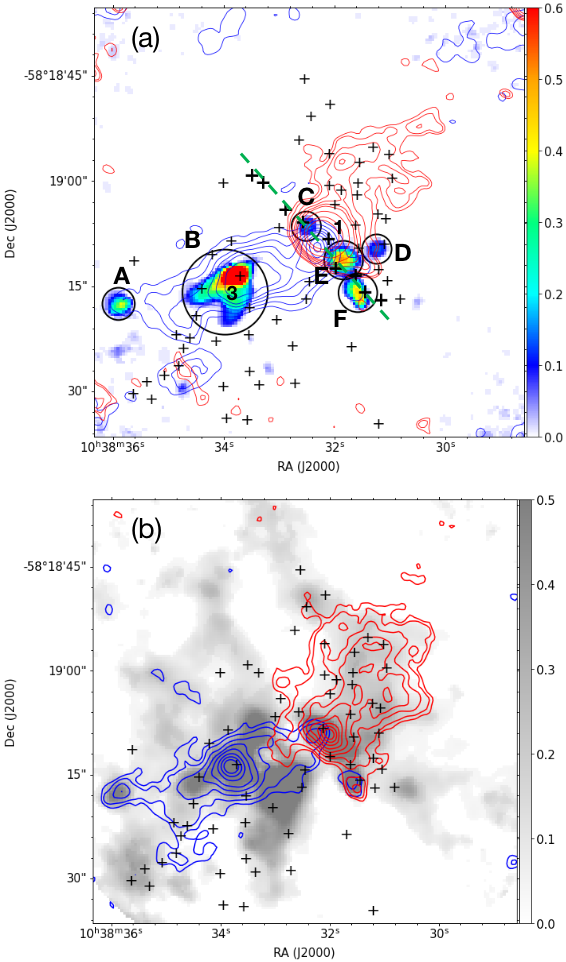}
\caption{(a) Contour of outflow traced by HCO$^{+}$ (1-0) superimposed on the moment 0 map of SiO (2-1), color bar on the right indicates the flux scale in Jy beam$^{-1}$ km/s. SiO emission only concentrate on several positions marked as A, B, C, D, E, F. Black circles represent the range of averaged spectra at each position. The SiO knot chain and core chain are marked by green dotted line. Contour levels for blue lobe are (0.2, 0.3, 0.5, 0.7, 0.9, 1.2, 1.5, 1.8) Jy beam$^{-1}$ km/s, (0.15, 0.2 to 1.1 by steps 0.1, 1.5, 2) Jy beam$^{-1}$ km/s for red lobe; (b) Contour of outflow traced by CS superimposed on the moment 0 map of SO. Contour levels for blue lobe are (0.15, 0.3, 0.5, 0.7, 0.9, 1.2, 1.5, 1.8) Jy beam$^{-1}$ km/s, (0.15, 0.2, 0.3, 0.4, 0.5, 0.6, 0.9, 1.5, 2) Jy beam$^{-1}$ km/s for red lobe. The velocity intervals for redshifted and blueshifted high velocity emission of HCO$^{+}$ (1-0) and CS (2-1) lines are defined as [-16.5, 5] km s$^{-1}$ and [-41, -23] km s$^{-1}$, respectively.}
\label{outflow}
\end{figure}

\begin{figure*}
\centering
\includegraphics[scale=0.4]{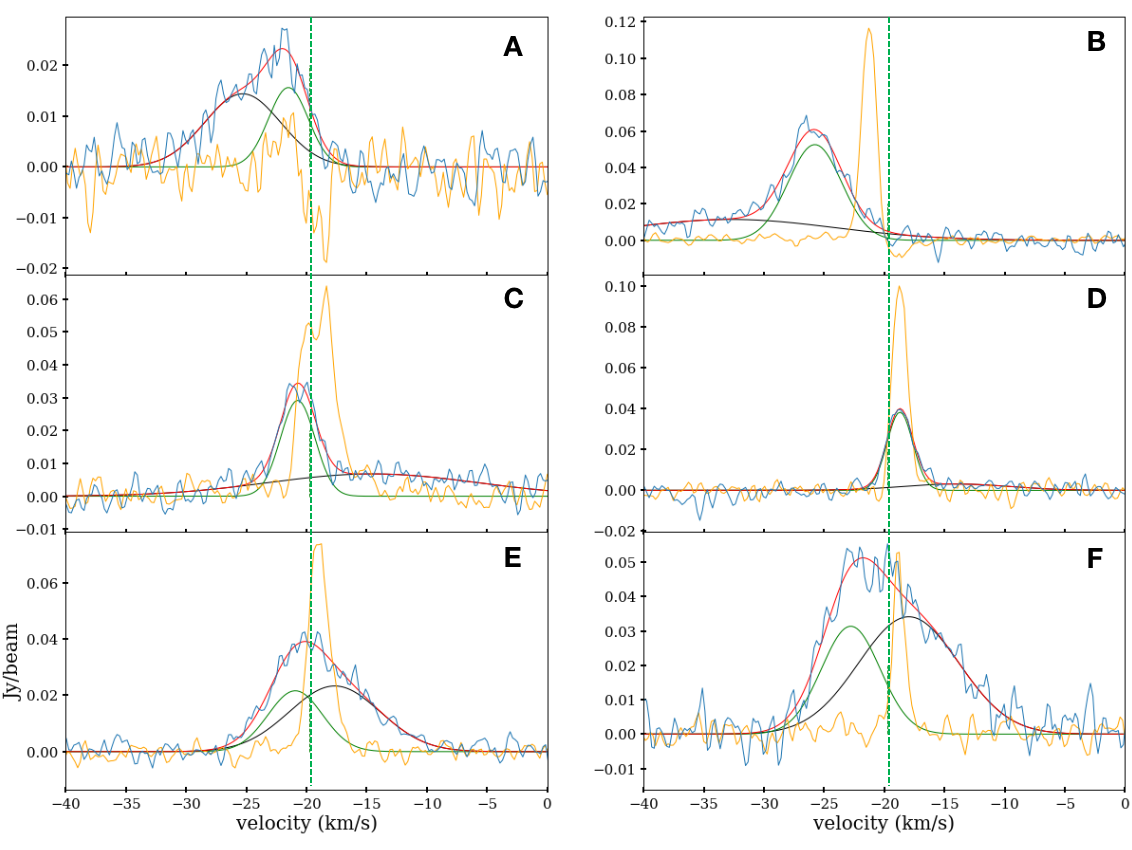}
\caption{Averaged spectra of six positions marked by black circles in Fig.~\ref{outflow}. The SiO (2-1) and H${^{13}}{\rm C}$O$^{+}$ (1-0) spectra are drawn in blue and orange lines, respectively. Black and green lines are to indicate the decomposed components of SiO(2-1) line and red spectral line is the result of double Gaussian fitting for SiO (2-1) line. Vertical green dotted line marks the systemic velocity.}
\label{width}
\end{figure*}

\begin{table}
	\centering
	\caption{Fitting parameters of SiO lines {\bf toward} G286.21+0.17. Peak intensity I$_P$, velocity V$_p$ corresponding to I$_P$ and line width FWHM. r is the radius of black circle representing the area over which the averaged spectra are taken at each position marked in Fig.~\ref{outflow}.}
	\label{wid}
	\begin{tabular}{ccccc} % five columns, alignment for each
		\hline
		Position   &r  & $I_{p}$  & FWHM  & $V_{p}$ \\
		        &(arcsecond)  & (Jy~beam$^{-1}$)  & (km~s$^{-1}$)  & (km~s$^{-1}$) \\
		\hline
A   &2.30  &0.032    & 3.74   &-21.73\\
         &       &0.018    & 8.01   & -25.21 \\                
B    &6.06  &0.021    & 8.11   &-25.75\\
         &       &0.007     & 19.69  & -36.44\\
C   &2.10  &0.028    & 3.43   &-20.76\\
         &       &0.005     & 20.25  &-15.32 \\
D   &2.10  &0.031    & 2.65   &-18.71\\
E   &2.76  &0.033    & 6.04   &-20.82\\
         &       &0.017    & 9.06   & -17.35 \\
F   &2.76  &0.027    & 5.49   &-22.82\\
         &       &0.019    & 10.23  & -18.67\\
		\hline

% \multicolumn{5}{l}{$^a$ Peak intensity I$_P$, velocity V$_p$ corresponding to I$_P$ and line width FWHM. r is the radius of black circle representing the area over which the averaged spectra are taken at each position marked in Fig.~\ref{outflow}.}\\
	\end{tabular}
\end{table}

\subsection{Molecular Outflows and shocked gas} \label{3.5}

As shown in Fig.~\ref{single}(c) and Fig.~\ref{blue127}, the spectral lines of optically thick HCO$^{+}$ (1-0) and CS (2-1) have obvious high-velocity line wings, strongly suggesting the existence of energetic outflows. Considering that most of the optically thin H$^{13}$CO$^+$ (1-0) line emission is between -23 km s$^{-1}$ and -16.5 km s$^{-1}$, the velocity intervals for redshifted and blueshifted high velocity
emission of HCO$^{+}$ (1-0) and CS (2-1) lines are defined as [-16.5, 5] km s$^{-1}$ and [-41, -23] km s$^{-1}$, respectively. The integrated intensity maps of outflow emission (or outflow lobes) are shown in Fig.~\ref{outflow}. The outflow seems to be driven by G286c1 (marked as "1"  in Fig.~\ref{hc3n}a or Fig.~\ref{fildc}). The outflow is well aligned with the skeleton of the longest sub-filament. In addition, the cavities in the moment 0 map of HC$_{3}$N (11-10) and also in the 3mm emission continuum (see Fig.~\ref{hc3n}a) are approximately along the blueshifted outflow lobe. Naturally, we can infer that the outflows swept gas out, and the mechanical force of outflows could contribute to the formation of the cavity.

As shown in Fig.~\ref{outflow}a, SiO (2-1) emission reveals several bright knots along the outflows, which are marked from "A" to "F". Knots A and B are located within the blueshifted lobe of outflow and others (Knots C, D, E, F) within the redshifted lobe. Almost every SiO emission knot has a corresponding dense core. Fig.~\ref{width} present averaged SiO spectra over the area defined by black circle in Fig.~\ref{outflow}a. The spectra can be well fitted with double gaussian profiles (a narrow component plus a broad component). The Gaussian fitting results are summarized in Tab.~\ref{wid}. The line widths of broad components in SiO emission are much larger than 8 km~s$^{-1}$, also suggesting they are affected by outflow shocks.

Besides the broad components, SiO lines also show a narrow component. Narrow SiO (2-1) emission has been detected in infrared dark clouds \citep{Jimenez2010,Csengeri2016,Cosentino2018,Cosentino2020} and high-mass proto-cluster forming regions \citep{Nguyen2013,Louvet2016,Csengeri2016,Liu2020}. Such narrow gaussian component may come from the thermal radiation of cores in molecular cloud, or reproduced by low-velocity shocks \citep{Louvet2016}. And low-velocity shocks are usually attributed to colliding flows or cloud-cloud collision \citep{Jimenez2010,Nguyen2013,Louvet2016,Liu2020}.
As shown in Fig.~\ref{width}, the narrow components of SiO (2-1) emission lines at knots C, E and F (the knot chain marked by green dashed line in Fig.~\ref{outflow}) are blueshifted with respected to the H$^{13}$CO$^+$ J=1-0 emission, although these knots are located within the redshifted outflow lobe. It is contrary to the broad components which are redshifted with respect to the systemic velocity, indicating that the narrow components of SiO (2-1) emission lines at these knots may not be likely produced by outflow. The SiO J=2-1 emission at knot D has smallest line width of 2.65 km~s$^{-1}$ and is peaked at similar velocity as H$^{13}$CO$^+$ J=1-0 emission. The SiO J=2-1 emission at knot D does not show clear high-velocity wing or broad emission, and thus it is not affected by outflows. As discussed in Sec.~\ref{4.3}, the SiO knots C, D, E, and F may originate from a collision scenario. However, we cannot fully confirm this scenario with present data.

As shown in Fig.~\ref{outflow}b, the emission of SO (3(2)-2(1)) is also strong at the positions of SiO knots, indicating that SO emission is also strengthened by outflow shock. Both the SiO (2-1) emission and SO (3(2)-2(1)) emission
are strongest at the position of G286c3 (marked as "3"  in Fig.~\ref{hc3n}a or Fig.~\ref{fildc}), and the bow-shaped structures from the moment 0 maps of HCO$^{+}$ (1-0) and CS (2-1) show in Fig.~\ref{moment127}. These evidences suggest that the formation of G286c3 is possibly triggered by the outflow driven by G286c1. 

\section{Discussion}

\subsection{Is G286 undergoing global collapse?}\label{4.1}

According to the model of contracting clouds (e.g. \citealt{Leung1977,Zhou1993,Myers1996}), a pair of optically thick and thin lines can be used for identifying infall motions. The optically thin line should show a single-peaked profile, which helps to rule out the possibilities of other gas kinematics (e.g., rotation or multiple velocity components). In contract, the optically thick lines of contracting clouds often show double-peaked profiles with blueshifted one stronger than the redshifted one. Such "blue profile" can be well produced by radiation transfer models for contracting clouds with density and temperature gradients (e.g. \citealt{Zhou1993,Myers1996,Lee1999}).

B10 argued that G286 is undergoing global collapse through analyzing the J=1-0 and J=4-3 transitions of HCO$^+$ and H$^{13}$CO$^+$ lines obtained in single dish observations. However, the optically thin line H${^{13}}{\rm C}$O$^{+}$ is very weak and fuzzy in Fig.4 of B10 due to the low resolution and sensitivity in their single-dish observations. They argued that H${^{13}}{\rm C}$O$^{+}$ line is single peaked and regarded it as an important evidence to testify the existence of global infall in the G286 region. However, the results of our higher resolution ALMA array data show the profile of optically thin line H${^{13}}{\rm C}$O$^{+}$ (1-0) is double-peak in Fig.~\ref{blue127} and Fig.~\ref{single}(c). In Fig.2 of C20, they presented spectra of CO (2-1), $\rm{C^{18}O}$ (2-1), $\rm{N_2D}^+$ (3–2) and DCO$^{+}$ (3–2) averaged over an area with $2.5\arcmin$ in radius centered on the phase center. The line profile of optically thick CO is single-peaked, but the line profiles of optically thin molecules are double-peaked. So the line profiles of C20 are different from the profiles by B10 and our study. C20 also analyzed the spectral-line profiles of individual cores (Appendix B of C20), they found that both the optically thin and optically thick molecular spectral-lines of G286c3 show double-peak profiles, and they attribute this to two different velocity components. In our ALMA observations, the optically thick and thin lines (such as HCO$^+$ (1-0), H${^{13}}{\rm C}$O$^{+}$ (1-0) and CCH (1-0)) also show double peaked profiles.
As discussed in the previous sections, the double peaked profiles are caused by other bulk motions (e.g., relative motions of two sub-clumps, outflows) as well as global infall within the clumps.

In short, infall motion in distant massive clumps should not be the only cause of producing blue profile of molecular lines in single dish observations. Thus, we advise great caution in interpreting gas kinematics (e.g., infall) from line profiles toward distant massive clumps in single dish observations. It is simply because that the low resolution single-dish observations can not resolve the complex gas motions within distant massive clumps \citep{Wu2003,Fuller2005,He2015, Yue2021}.

Although the "blue profile" of optically thick line emission in single dish observations of G286 (like B10) is not likely caused by gas infall motions, this cannot rule out the possibility that G286 is still undergoing overall collapse. Following \citet{Wuj2010}, the virial mass of the dense clump is calculated by
\begin{equation}
M_{Vir}=\frac{5R\Delta v^{2}}{8a_{1}a_{2}Gln2}
 \sim 209\frac{(R/1 pc)(\Delta v/1 km/s)}{a_{1}a_{2}} M_{\odot}
\end{equation}
\begin{equation}
, a_{1}=\frac{1-p/3}{1-2p/5}, \ p < 2.5.
\end{equation}
where $a_{1}$ is the correction for a power-law density distribution ($n(r) \propto r^{-p}$) and
$a_{2}$ is the correction for a nonspherical shape (\citealt{Bertoldi1992}). In Sec.~\ref{3.1}, we assume the spherical geometry of clump, $a_{2}= 1$. To calculate $a_{1}$, we take the average value $p \sim 1.77$ \citep{Mueller2002} for all massive dense clumps. The linewidth of H${^{13}}{\rm C}$O$^{+}$ J=1-0 is $\Delta v \sim 1.46$ km~s$^{-1}$ by fitting the ACA 7m spectral line (see Fig.~\ref{single}(c)) with two Gaussian profiles. The radius R derived from H$^{13}$CO$^{+}$ J=1-0 emission is 0.40 $\pm$ 0.05 pc (see B10). Then the virial mass of G286 clump is $M_{Vir} \sim 126~$M$_{\odot}$, which is significantly smaller than the total mass of G286 clump. This indicates that the G286 clump is still bounded by gravity and may be undergoing overall collapse. As discussed in next section, the energetic outflow cannot push the two sub-clumps away. In addition, G286 is also confined by external pressure from nearby H{\sc ii} regions. Therefore, the G286 clump will likely collapse further, and the two sub-clumps will move close to each other.

\subsection{Feedback of outflows}\label{4.2}

Energetic high-velocity outflows in massive protoclusters could entrain wide-angle low-velocity materials through momentum feedback \citep{Liu2010, Liu2016b, Bally2016}. Below we discuss whether the energetic outflows in G286 can drive expansion of the two sub-clumps. 

The mass ($M_{\rm gas}$), momentum ($P_{\rm gas}$) and energy ($E_{\rm gas}$) of the two sub-clumps can be estimated with line emission of H$^{13}$CO$^+$ J=1-0. The mass ($M_{\rm out}$), momentum ($P_{\rm out}$) and energy ($E_{\rm out}$) of the outflows can be estimated from the line emission of high velocity wings in HCO$^+$ J=1-0 line. We assume that the line emission of H$^{13}$CO$^+$ J=1-0 and high velocity emission of HCO$^+$ J=1-0 are optically thin.

Following the approach described in \citet{Li2019}, we estimated the physical parameters of the outflows and the two sub-clumps as below. We assume that the excitation temperature of H$^{13}$CO$^+$ J=1-0 and HCO$^+$ J=1-0 lines are the same and constant in both low-velocity gas and high-velocity outflow gas. Therefore, the masses are: 
\begin{equation}
\label{Mout}
M_{\rm out} \propto \left[\frac{\rm H_{2}}{\rm HCO^{+}}\right] 
\int_{\Omega} N_{\rm HCO^{+}}(\Omega) d\Omega \\
\propto \left[\frac{\rm H_{2}}{\rm HCO^{+}}\right] \int_{\Omega} \int_{v} S_{v}dv d\Omega.
\end{equation}

\begin{equation}
\label{Mout}
M_{\rm gas} \propto \left[\frac{\rm H_{2}}{\rm H^{13}CO^{+}}\right] \int_{\Omega} \int_{v^{'}} S_{v^{'}}dv^{'} d\Omega,
\end{equation}

The expression in bracket is the abundance ratio of molecules. The momentum of outflows and G286 sub-clump are:
\begin{equation}
\label{Pout}
P_{\rm out} = M^{r}_{\rm out} v^{r}_{out} + M^{b}_{out} v^{b}_{\rm out},
\end{equation}
\begin{equation}
\label{Pgas}
P_{\rm gas} = M^{r}_{\rm gas} v^{r}_{gas} + M^{b}_{gas} v^{b}_{\rm gas}.
\end{equation}
From above formulas, we get the momentum ratio:
\begin{equation}
\label{ratio1}
\frac{P_{\rm out}}{P_{\rm gas}}\approx \left[\frac{\rm H^{13}CO^{+}}{\rm HCO^{+}}\right]\frac{S^{r}_{\rm HCO^{+}} v^{r}_{out} + S^{b}_{\rm HCO^{+}} v^{b}_{out}}{S^{r}_{\rm H^{13}CO^{+}} v^{r}_{out} + S^{b}_{\rm H^{13}CO^{+}} v^{b}_{out}}.
\end{equation} 

Here v$^{r}$ and v$^{b}$ are the characteristic velocities of redshifted and blueshifted gas. S$^{r}$ and S$^{b}$
are the corresponding total flux. The total flux can be estimated from the moment 0 maps of HCO$^{+}$ and H$^{13}$CO$^{+}$. The redshifted velocity v$^{r}$ and blueshifted velocity v$^{b}$ are defined as the median velocities of outflow lobes or sub-clumps in the moment 1 maps. After subtracting the systemic velocity, the median intensity-weighted velocities of the redshifted and blueshifted  outflows are $v_{out}^r\sim 3.5$ km s$^{-1}$ and $v_{out}^b\sim 3.8$ km s$^{-1}$, respectively.  Similarly, we obtained $v_{gas}^r\sim 1.0$ km s$^{-1}$ for the redshifted sub-clump and $v_{gas}^b\sim 1.1$ km s$^{-1}$ for the blueshifted sub-clump. Taking $\left[ \rm H^{13}CO^{+}/ \rm HCO^{+}\right]\sim 1/60$ \citep{Milam2005}, the momentum ratio is $P_{\rm out}/P_{\rm gas}\sim 0.07$. Although there are many uncertainties (e.g., excitation temperature, optical depths and abundances) in above analysis, the momentum of the outflows seems much smaller than that of the two gas clumps.

Further we can calculate the ratio of energy:
\begin{equation}
\label{Eout}
E_{\rm out} = \frac{1}{2}M^{r}_{\rm out} (v^{r}_{out})^2 + \frac{1}{2}M^{b}_{\rm out} (v^{b}_{out})^2,
\end{equation}
\begin{equation}
\label{Pgas}
E_{\rm gas} = \frac{1}{2}M^{r}_{\rm gas} (v^{r}_{gas})^2 + \frac{1}{2}M^{b}_{\rm gas} (v^{b}_{gas})^2,
\end{equation}
then we obtain $E_{\rm out}/E_{\rm gas}\sim 0.24$. The momentum and energy of outflows in G286 are significantly smaller than those of the two gas sub-clumps, indicating that the outflows are not capable of driving the two sub-clumps.

The dynamical time of outflow can be calculated by 
\begin{equation}
\label{tdyn}
t_{\rm dyn} = \frac{\lambda_{\rm max}}{(v^{b}_{\rm max} + v^{r}_{\rm max})/2}.
\end{equation}
The length of longer blue lobe ($\lambda_{\rm max}$) is $\sim 0.40$ pc. $v^{r}_{\rm max}$ and $v^{b}_{\rm max}$ are the maximum velocities of redshifted and blueshifted  outflow lobes, respectively.  We obtain $t_{\rm dyn} \sim 1.13 \times 10^{4}$ yr, which is consistent with the fact that G286 is a very young cluster. However, here we don't consider the projection effect. From above analysis, we argue that the outflow feedback in G286 is not sufficient to drive expansion of the two sub-clumps or to cause the large-scale velocity gradients across the whole G286 region. In particular, outflows cannot explain the large-scale velocity gradient along the south-west to north-east direction (see Fig.~\ref{big}b ). In addition, there are no outflows in the northern region close to the H{\sc ii} region A. Therefore, velocity gradients in regions adjacent to H{\sc ii} region A (see Fig.~\ref{big}c) are more likely caused by the expansion of H{\sc ii} region rather than outflows.

\subsection{Formation of the "L" type filament and the G286 protocluster}\label{4.3}

\begin{figure*}
\centering
\includegraphics[scale=1.4]{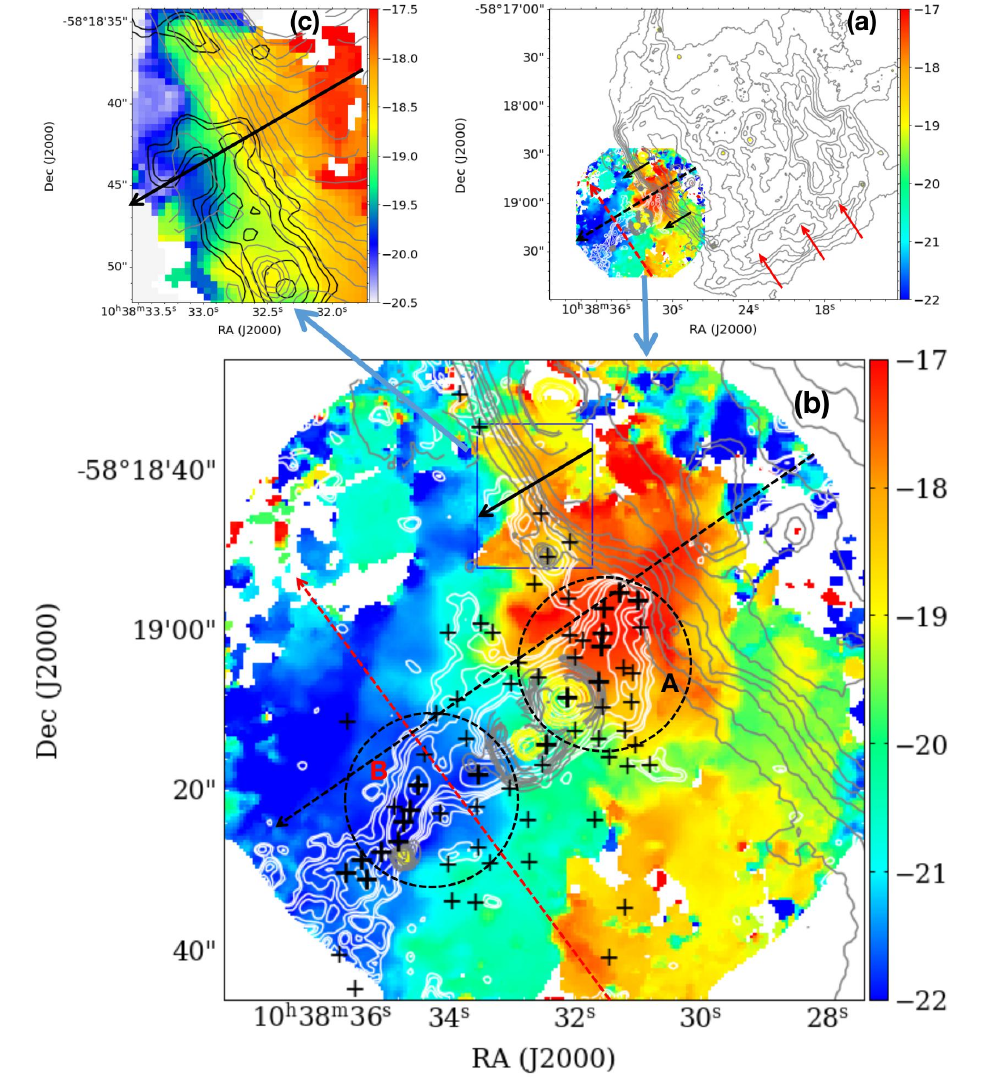}
\caption{(a) Lower left is the velocity field in G286 region traced by HCO$^{+}$ (1-0), here we reduce the outflow (as Fig.~\ref{moment127} shows). White contour represents the emission of 3 mm continuum. Grey contours (Spitzer 8 $\mu$m emission) show the H{\sc ii} region A. Black arrows indicate the approximate extrusion direction of H{\sc ii} region A, red arrows for H{\sc ii} region B. Dashed black arrow and red arrow indicate the velocity gradient from extrusion. Contour levels for 3mm continuum are 1$\sigma$ $\times$ (1.5, 2, 3, 4, 5, 8, 12, 16, 24, 32, 48, 64, 75) with $\sigma$= 0.21 mJy/beam; (b) 7 m+12 m moment 1 map of HCO$^{+}$ for G286 region, for eliminating outflow,  velocity range is limited as (-23 km/s, -16.5 km/s), the systemic velocity is -19.8 km/s. Yellow contours (Spitzer 4.5 $\mu$m emission) show two protostars G286c1 and G286c2, others are the same as (a). Contour levels for Spitzer 4.5 $\mu$m and 8 $\mu$m are (80, 100, 120, 140, 160, 200, 280, 360, 390) mJy/sr. Dashed black circles roughly mark two sub-clumps; (c) Velocity field of CCH (1-0), the size of region is the blue box marked in (b).}
\label{big}
\end{figure*}

\begin{figure*}
\centering
\includegraphics[scale=0.28]{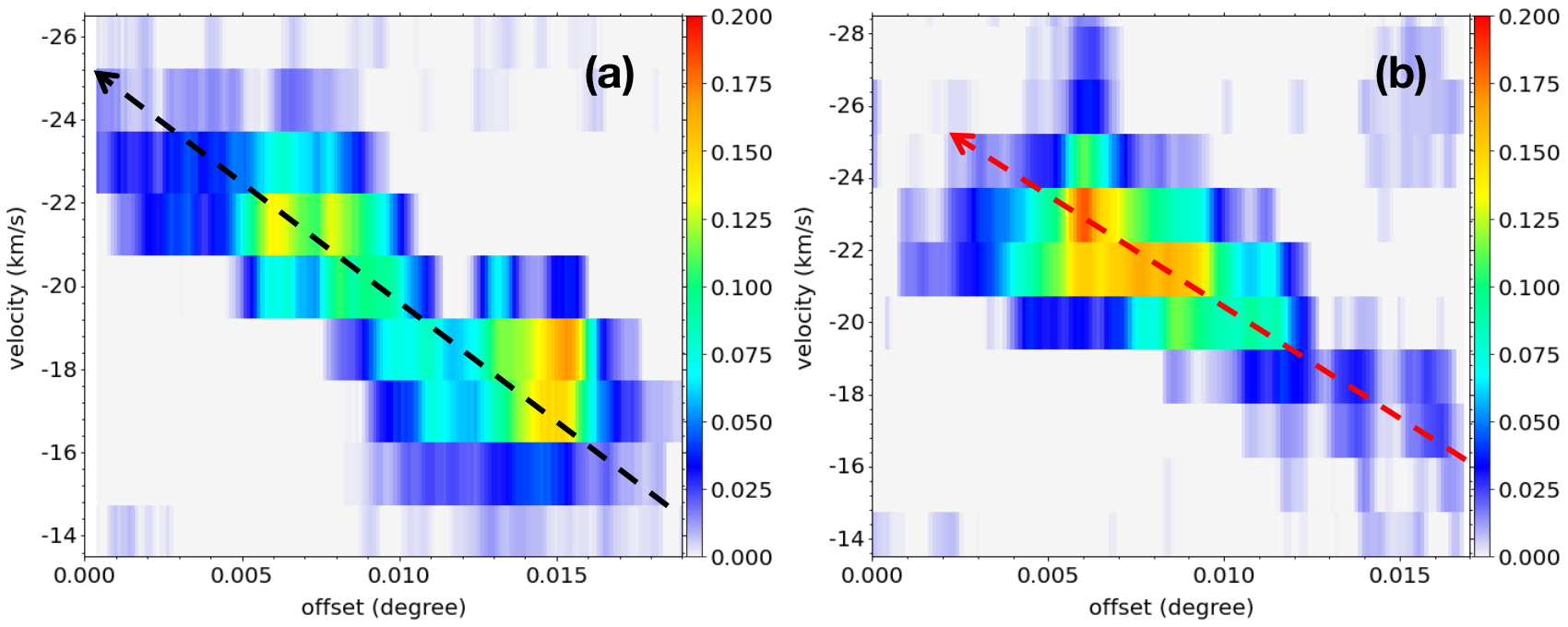}
\caption{PV diagram of CS J=2-1 along the black and red dashed arrows in Fig.~\ref{big}(a).}
\label{pv}
\end{figure*}

% \begin{figure*}
% \centering
% \includegraphics[scale=0.5]{fig/big2.png}
% \caption{}
% \label{big2}
% \end{figure*}

As shown in Fig.~\ref{big}(a), the H{\sc ii} region A seems to strongly interact with the G286 clump. The intensity gradient of 8 µm emission associated with H{\sc ii} region A is found to increase toward the G286 clump, suggesting that the H{\sc ii} region is compressing the molecular gas of G286 clump. Fig.~\ref{big}(b) presents the moment 1 maps of HCO$^+$ J=1-0. Overall, there is a clear velocity gradient along the north-west to south-east direction. Fig.~\ref{pv}(a) shows the position-velocity (PV) diagram of CS J=2-1 along the black dashed arrow in Fig.~\ref{big}(a). The velocity gradient inferred from this PV diagram is about 14 km~s$^{-1}$~pc$^{-1}$. In addition, we also identified a velocity gradient along the south-west to north-east direction, which is across the “NW-SE” filament of G286. Fig.~\ref{pv}(b) presents the position-velocity (PV) diagram of CS J=2-1 along the red dashed arrow in Fig.~\ref{big}(a), which reveals a velocity gradient of about 13 km~s$^{-1}$~pc$^{-1}$. We also display the PV diagrams of HCO$^+$ J=1-0, H$^{13}$CO$^+$ J=1-0, and CCH J=1-0 for both directions mentioned above in Appendix.~\ref{C}, and all of those PV diagrams show distinct velocity gradients similar to that of CS J=2-1 in Fig.~\ref{pv}. 

Filaments with prominent velocity gradients
perpendicular to their major axes have been detected in molecular clouds \citep{Beuther2015,Chen2020}. Such velocity gradients across filaments suggest that self-gravitating filaments may form from shocked flows in molecular clouds \citep{Chen2020}. The velocity gradient in the G286 clump along south-west to north-east direction could be caused by such large-scale compression flows. The H{\sc ii} region A seems also to be compressed by the same large-scale flow. As indicated by the red arrows in Fig.~\ref{big}(a), the contours in the southern part of H{\sc ii} region A  are much denser than the contours in the northern part, indicating that the southern part may be compressed or confined by external pressure. As shown in Fig.~\ref{red}(a), both the G286 clump and the H{\sc ii} region A are located to the north of a giant H{\sc ii} region B (Gum31) that is excited by a massive star cluster NGC3324. They seemingly locate at the loophole of H{\sc ii} region B. The gas blowing from H{\sc ii} region B seems to be acting on them. As seen in Fig.5 of \citet{Rebolledo2016}, there is a clear velocity gradient due to the expansion of H{\sc ii} region B. Both G286 clump and H{\sc ii} region A are located along that large scale velocity gradient. Therefore, we suspect that the expansion of H{\sc ii} region B compresses its surrounding interstellar medium and induces the large-scale velocity gradient across the “NW-SE” filament within the G286 clump. More discussions on the stellar feedback of H{\sc ii} region B are presented in Appendix.~\ref{C}. However, we cannot rule out the possibility that the large-scale velocity gradient along south-west to north-east direction is caused by large-scale shocked flows induced by supersonic turbulence \citep{Chen2020}.

The northen part of the G286 clump, however, is more affected by the expansion of H{\sc ii} region A. Fig.~\ref{big}(c) shows the moment 1 map of CCH J=1-0 toward the "NE-SW" filament. CCH is a good tracer of low density PDR regions \citep{Cuadrado2015}. As shown in Fig.~\ref{big}(c), there is an obvious velocity gradient across the "NE-SW" filament, which is likely caused by the expansion of the PDR of H{\sc ii} region A. Moreover, the 3 mm continuum emission ("NE-SW" filament) is just located in front of the PDR region, the protruding portion of 3 mm continuum emission just corresponds to the sagging portion of H{\sc ii} region A (right upper edge of sub-clump A in Fig.~\ref{big}(b)), which also indicates that “NE-SW” filament is created by the extrusion of H{\sc ii} region A. 
At the middle region of Fig.~\ref{big}(b), there is a core chain centered on G286c1 at the interface of the “NW-SE” and “NE-SW” filaments. A chain of SiO knots showing narrow line emission is also along this core chain (as shown in the green dashed line in Fig.~\ref{outflow}(a)), which may indicate a shock front induced by the collision of the “NW-SE” and “NE-SW” filaments.
\begin{figure*}
\centering
\includegraphics[scale=0.45]{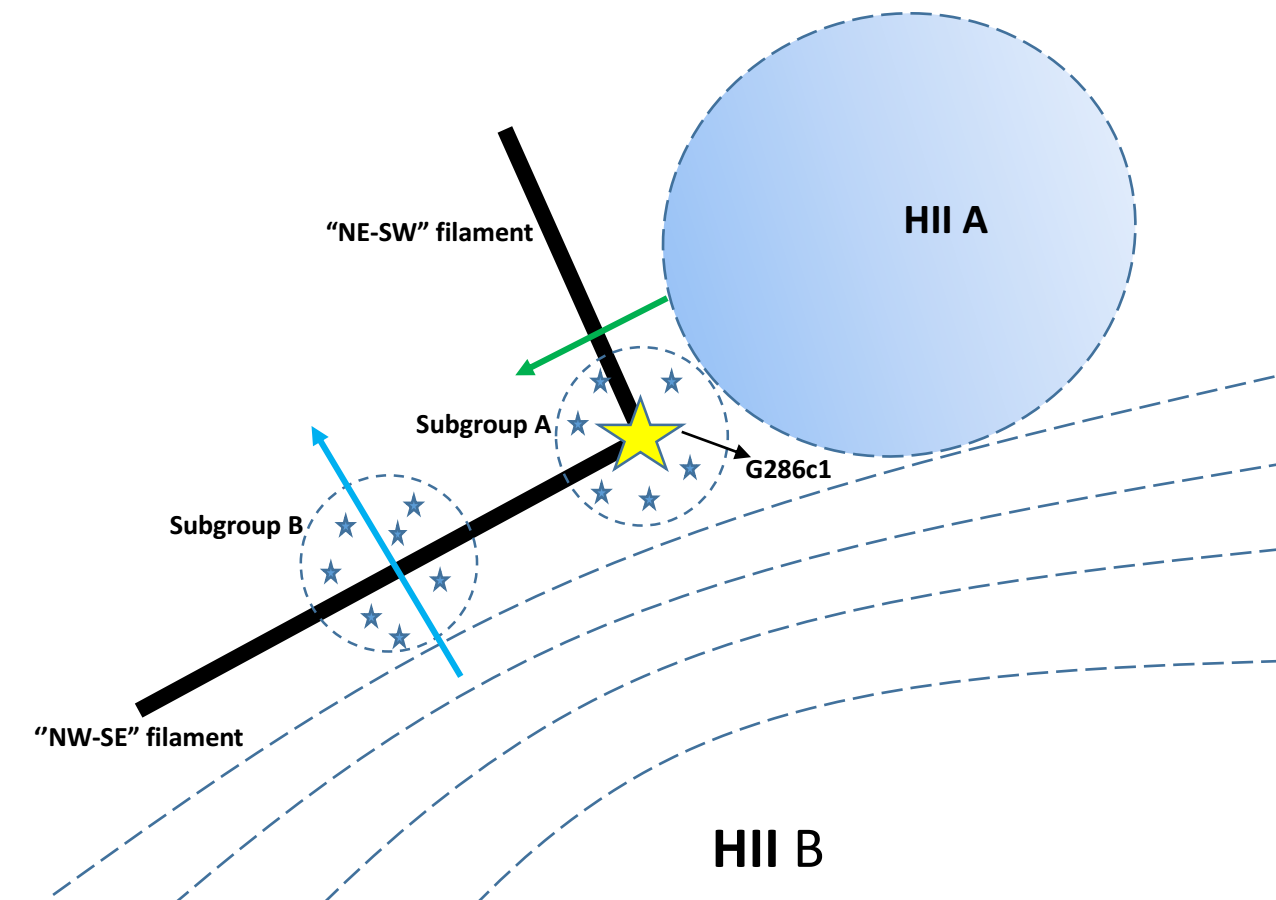}
\caption{Schematic diagram of the star formation history in G286 region.}
\label{map}
\end{figure*}

Fig.~\ref{map} presents a schematic cartoon that summarizes the key points of the formation process of the "L" type filaments and proto-clusters in the G286 clump:

(1) Two H{\sc ii} regions are squeezing the G286 clump. The green and cyan arrows indicate the directions of extrusion by the two H{\sc ii} regions and the velocity gradients induced by the compression flows. 

(2) Two main filaments (“NW-SE” and “NE-SW” filaments) with prominent velocity gradients
perpendicular to their major axes are formed due to the compression flows. The collision of the “NW-SE” and “NE-SW” filaments generate the ''L'' type structure.

(3) The filaments fragment into two subgroups of dense cores that are forming the cluster of stars. The most massive core G286c1 forms at the junction of “NE-SW” and “NW-SE” filaments. G286c1 is driving an energetic molecular outflow, which reshapes the gas distribution and creates a cavity in the “NW-SE” filament, further triggers the formation of strongly shocked core G286c3.

\section{Summary}
We have studied the gas kinematics in G286 clump from large scale to small scale with ALMA data obtained in the "ATOMS" project. The main results are as follows:

(1) The total mass of the central region of G286 clump is estimated as M $\sim$ 1021 M$_{\sun}$. The skeletons of sub-filaments in G286 are identified by the FILFINDER algorithm, and the most massive compact cores identified by the Dendrogram algorithm are located along the longest sub-filament with a length of $\sim0.77$ pc. The width of longest filament is W$\sim$0.075 pc. We obtain two subgroups of dense cores from a nearest neighbor separation (NNS) analysis, which are perfectly matched to the "red" group and "blue" group of C20. 

(2) We demonstrate that the blue profile of molecular lines in previous studies (B10) of G286 with single dishes is actually caused by gas emission from two sub-clumps rather than gas infall. We advise great caution in interpreting gas kinematics (e.g., infall) from line profiles toward distant massive clumps in single dish observations because the low resolution single-dish
observations can not resolve their complex gas motions inside.

(4) Energetic outflows are identified in G286 clump with high velocity emission in HCO$^{+}$ J=1-0 and CS J=2-1 lines. The outflow seems to be driven by G286c1 and it well aligns with the skeleton of filament. The outflow has created a cavity in G286 clump but it is not strong enough to drive expansion of the two sub-clumps.

(5) The two main filaments (“NW-SE” and “NE-SW” filaments) show prominent velocity gradients perpendicular to their major axes. The “NE-SW” filament seems to be formed due to the compression by the PDR of H{\sc ii} region A. We also identify a velocity gradient along the south-west to north-east direction across the “NW-SE” filament, which could be caused by the large-scale compression flows induced by the expansion of giant HII region B. The collision of the “NW-SE” and “NE-SW” filaments generate the ''L'' type structure in G286. And the most massive core G286c1 forms at the junction point of the “L” structure. 

% \acknowledgments
% \bibliography{G286}{}
% \bibliographystyle{aasjournal}
% \end{document}

\section*{Acknowledgements}
Tie Liu acknowledges the supports by National Natural Science Foundation of China (NSFC) through grants No.12073061 and No.12122307, the international partnership program of Chinese academy of sciences through grant No.114231KYSB20200009, and Shanghai Pujiang Program 20PJ1415500. C.W.L. is supported by the Basic Science Research Program through the National Research Foundation of Korea (NRF) funded by the Ministry of Education, Science and Technology (NRF-2019R1A2C1010851).
Guo-Yin ZHANG acknowledges the supports by China Postdoctoral Science Foundation (No. 2021T140672).
%%%%%%%%%%%%%%%%%%%%%%%%%%%%%%%%%%%%%%%%%%%%%%%%%%
\section{Data availability}The data underlying this article are available in the article and in ALMA archive.


%\bibliographystyle{mnras}

% \begin{thebibliography}{99}
% \end{thebibliography}
\bibliographystyle{mnras}
\bibliography{G286}

\begin{thebibliography}{}
\makeatletter
\relax
\def\mn@urlcharsother{\let\do\@makeother \do\$\do\&\do\#\do\^\do\_\do\%\do\~}
\def\mn@doi{\begingroup\mn@urlcharsother \@ifnextchar [ {\mn@doi@}
  {\mn@doi@[]}}
\def\mn@doi@[#1]#2{\def\@tempa{#1}\ifx\@tempa\@empty \href
  {http://dx.doi.org/#2} {doi:#2}\else \href {http://dx.doi.org/#2} {#1}\fi
  \endgroup}
\def\mn@eprint#1#2{\mn@eprint@#1:#2::\@nil}
\def\mn@eprint@arXiv#1{\href {http://arxiv.org/abs/#1} {{\tt arXiv:#1}}}
\def\mn@eprint@dblp#1{\href {http://dblp.uni-trier.de/rec/bibtex/#1.xml}
  {dblp:#1}}
\def\mn@eprint@#1:#2:#3:#4\@nil{\def\@tempa {#1}\def\@tempb {#2}\def\@tempc
  {#3}\ifx \@tempc \@empty \let \@tempc \@tempb \let \@tempb \@tempa \fi \ifx
  \@tempb \@empty \def\@tempb {arXiv}\fi \@ifundefined
  {mn@eprint@\@tempb}{\@tempb:\@tempc}{\expandafter \expandafter \csname
  mn@eprint@\@tempb\endcsname \expandafter{\@tempc}}}

\bibitem[\protect\citeauthoryear{{Andersen}, {Barnes}, {Tan}, {Kainulainen}  \&
  {de Marchi}}{{Andersen} et~al.}{2017}]{Andersen2017}
{Andersen} M.,  {Barnes} P.~J.,  {Tan} J.~C.,  {Kainulainen} J.,   {de Marchi}
  G.,  2017, \mn@doi [\apj] {10.3847/1538-4357/aa9072}, \href
  {https://ui.adsabs.harvard.edu/abs/2017ApJ...850...12A} {850, 12}

\bibitem[\protect\citeauthoryear{{Andr{\'e}} et~al.,}{{Andr{\'e}}
  et~al.}{2010}]{Andre2010}
{Andr{\'e}} P.,  et~al., 2010, \mn@doi [\aap] {10.1051/0004-6361/201014666},
  \href {https://ui.adsabs.harvard.edu/abs/2010A&A...518L.102A} {518, L102}

\bibitem[\protect\citeauthoryear{{Andr{\'e}}, {Di Francesco}, {Ward-Thompson},
  {Inutsuka}, {Pudritz}  \& {Pineda}}{{Andr{\'e}} et~al.}{2014}]{Andre2014}
{Andr{\'e}} P.,  {Di Francesco} J.,  {Ward-Thompson} D.,  {Inutsuka} S.~I.,
  {Pudritz} R.~E.,   {Pineda} J.~E.,  2014, in {Beuther} H.,  {Klessen} R.~S.,
  {Dullemond} C.~P.,   {Henning} T.,  eds, Protostars and Planets VI. p.~27
  (\mn@eprint {arXiv} {1312.6232}),
  \mn@doi{10.2458/azu_uapress_9780816531240-ch002}

\bibitem[\protect\citeauthoryear{{Arzoumanian} et~al.,}{{Arzoumanian}
  et~al.}{2011}]{Arzoumanian2011}
{Arzoumanian} D.,  et~al., 2011, \mn@doi [\aap] {10.1051/0004-6361/201116596},
  \href {https://ui.adsabs.harvard.edu/abs/2011A&A...529L...6A} {529, L6}

\bibitem[\protect\citeauthoryear{{Arzoumanian} et~al.,}{{Arzoumanian}
  et~al.}{2019}]{Arzoumanian2019}
{Arzoumanian} D.,  et~al., 2019, \mn@doi [\aap] {10.1051/0004-6361/201832725},
  \href {https://ui.adsabs.harvard.edu/abs/2019A&A...621A..42A} {621, A42}

\bibitem[\protect\citeauthoryear{{Bally}}{{Bally}}{2016}]{Bally2016}
{Bally} J.,  2016, \mn@doi [\araa] {10.1146/annurev-astro-081915-023341}, \href
  {https://ui.adsabs.harvard.edu/abs/2016ARA&A..54..491B} {54, 491}

\bibitem[\protect\citeauthoryear{{Barnes}, {Yonekura}, {Ryder}, {Hopkins},
  {Miyamoto}, {Furukawa}  \& {Fukui}}{{Barnes} et~al.}{2010}]{Barnes10}
{Barnes} P.~J.,  {Yonekura} Y.,  {Ryder} S.~D.,  {Hopkins} A.~M.,  {Miyamoto}
  Y.,  {Furukawa} N.,   {Fukui} Y.,  2010, \mn@doi [\mnras]
  {10.1111/j.1365-2966.2009.15890.x}, \href
  {https://ui.adsabs.harvard.edu/abs/2010MNRAS.402...73B} {402, 73}

\bibitem[\protect\citeauthoryear{{Benjamin} et~al.,}{{Benjamin}
  et~al.}{2003}]{Benjamin2003}
{Benjamin} R.~A.,  et~al., 2003, \mn@doi [\pasp] {10.1086/376696}, \href
  {https://ui.adsabs.harvard.edu/abs/2003PASP..115..953B} {115, 953}

\bibitem[\protect\citeauthoryear{{Bertoldi} \& {McKee}}{{Bertoldi} \&
  {McKee}}{1992}]{Bertoldi1992}
{Bertoldi} F.,  {McKee} C.~F.,  1992, \mn@doi [\apj] {10.1086/171638}, \href
  {https://ui.adsabs.harvard.edu/abs/1992ApJ...395..140B} {395, 140}

\bibitem[\protect\citeauthoryear{{Beuther}, {Ragan}, {Johnston}, {Henning},
  {Hacar}  \& {Kainulainen}}{{Beuther} et~al.}{2015}]{Beuther2015}
{Beuther} H.,  {Ragan} S.~E.,  {Johnston} K.,  {Henning} T.,  {Hacar} A.,
  {Kainulainen} J.~T.,  2015, \mn@doi [\aap] {10.1051/0004-6361/201527108},
  \href {https://ui.adsabs.harvard.edu/abs/2015A&A...584A..67B} {584, A67}

\bibitem[\protect\citeauthoryear{{Bonnell}, {Bate}, {Clarke}  \&
  {Pringle}}{{Bonnell} et~al.}{1997}]{Bonnell1997}
{Bonnell} I.~A.,  {Bate} M.~R.,  {Clarke} C.~J.,   {Pringle} J.~E.,  1997,
  \mn@doi [\mnras] {10.1093/mnras/285.1.201}, \href
  {https://ui.adsabs.harvard.edu/abs/1997MNRAS.285..201B} {285, 201}

\bibitem[\protect\citeauthoryear{{Bonnell}, {Bate}, {Clarke}  \&
  {Pringle}}{{Bonnell} et~al.}{2001}]{Bonnell2001}
{Bonnell} I.~A.,  {Bate} M.~R.,  {Clarke} C.~J.,   {Pringle} J.~E.,  2001,
  \mn@doi [\mnras] {10.1046/j.1365-8711.2001.04270.x}, \href
  {https://ui.adsabs.harvard.edu/abs/2001MNRAS.323..785B} {323, 785}

\bibitem[\protect\citeauthoryear{{Cappa}, {Niemela}, {Amor{\'\i}n}  \&
  {Vasquez}}{{Cappa} et~al.}{2008}]{Cappa2008}
{Cappa} C.,  {Niemela} V.~S.,  {Amor{\'\i}n} R.,   {Vasquez} J.,  2008, \mn@doi
  [\aap] {10.1051/0004-6361:20067028}, \href
  {https://ui.adsabs.harvard.edu/abs/2008A&A...477..173C} {477, 173}

\bibitem[\protect\citeauthoryear{{Carey} et~al.,}{{Carey}
  et~al.}{2009}]{Carey2009}
{Carey} S.~J.,  et~al., 2009, \mn@doi [\pasp] {10.1086/596581}, \href
  {https://ui.adsabs.harvard.edu/abs/2009PASP..121...76C} {121, 76}

\bibitem[\protect\citeauthoryear{{Chen}, {Mundy}, {Ostriker}, {Storm}  \&
  {Dhabal}}{{Chen} et~al.}{2020}]{Chen2020}
{Chen} C.-Y.,  {Mundy} L.~G.,  {Ostriker} E.~C.,  {Storm} S.,   {Dhabal} A.,
  2020, \mn@doi [\mnras] {10.1093/mnras/staa960}, \href
  {https://ui.adsabs.harvard.edu/abs/2020MNRAS.494.3675C} {494, 3675}

\bibitem[\protect\citeauthoryear{{Cheng}, {Tan}, {Liu}, {Kong}, {Lim},
  {Andersen}  \& {Da Rio}}{{Cheng} et~al.}{2018}]{Cheng2018}
{Cheng} Y.,  {Tan} J.~C.,  {Liu} M.,  {Kong} S.,  {Lim} W.,  {Andersen} M.,
  {Da Rio} N.,  2018, \mn@doi [\apj] {10.3847/1538-4357/aaa3f1}, \href
  {https://ui.adsabs.harvard.edu/abs/2018ApJ...853..160C} {853, 160}

\bibitem[\protect\citeauthoryear{{Cheng}, {Tan}, {Liu}, {Lim}  \&
  {Andersen}}{{Cheng} et~al.}{2020a}]{Cheng20}
{Cheng} Y.,  {Tan} J.~C.,  {Liu} M.,  {Lim} W.,   {Andersen} M.,  2020a,
  \mn@doi [\apj] {10.3847/1538-4357/ab879f}, \href
  {https://ui.adsabs.harvard.edu/abs/2020ApJ...894...87C} {894, 87}

\bibitem[\protect\citeauthoryear{{Cheng}, {Andersen}  \& {Tan}}{{Cheng}
  et~al.}{2020b}]{Cheng2020ApJ}
{Cheng} Y.,  {Andersen} M.,   {Tan} J.,  2020b, \mn@doi [\apj]
  {10.3847/1538-4357/ab93bc}, \href
  {https://ui.adsabs.harvard.edu/abs/2020ApJ...897...51C} {897, 51}

\bibitem[\protect\citeauthoryear{{Cosentino} et~al.,}{{Cosentino}
  et~al.}{2018}]{Cosentino2018}
{Cosentino} G.,  et~al., 2018, \mn@doi [\mnras] {10.1093/mnras/stx3013}, \href
  {https://ui.adsabs.harvard.edu/abs/2018MNRAS.474.3760C} {474, 3760}

\bibitem[\protect\citeauthoryear{{Cosentino} et~al.,}{{Cosentino}
  et~al.}{2020}]{Cosentino2020}
{Cosentino} G.,  et~al., 2020, \mn@doi [\mnras] {10.1093/mnras/staa2942}, \href
  {https://ui.adsabs.harvard.edu/abs/2020MNRAS.499.1666C} {499, 1666}

\bibitem[\protect\citeauthoryear{{Csengeri} et~al.,}{{Csengeri}
  et~al.}{2016a}]{Csengeri16}
{Csengeri} T.,  et~al., 2016a, \mn@doi [\aap] {10.1051/0004-6361/201526639},
  \href {https://ui.adsabs.harvard.edu/abs/2016A&A...585A.104C} {585, A104}

\bibitem[\protect\citeauthoryear{{Csengeri} et~al.,}{{Csengeri}
  et~al.}{2016b}]{Csengeri2016}
{Csengeri} T.,  et~al., 2016b, \mn@doi [\aap] {10.1051/0004-6361/201425404},
  \href {https://ui.adsabs.harvard.edu/abs/2016A&A...586A.149C} {586, A149}

\bibitem[\protect\citeauthoryear{{Cuadrado}, {Goicoechea}, {Pilleri},
  {Cernicharo}, {Fuente}  \& {Joblin}}{{Cuadrado} et~al.}{2015}]{Cuadrado2015}
{Cuadrado} S.,  {Goicoechea} J.~R.,  {Pilleri} P.,  {Cernicharo} J.,  {Fuente}
  A.,   {Joblin} C.,  2015, \mn@doi [\aap] {10.1051/0004-6361/201424568}, \href
  {https://ui.adsabs.harvard.edu/abs/2015A&A...575A..82C} {575, A82}

\bibitem[\protect\citeauthoryear{{Dame}, {Hartmann}  \& {Thaddeus}}{{Dame}
  et~al.}{2001}]{Dame2001}
{Dame} T.~M.,  {Hartmann} D.,   {Thaddeus} P.,  2001, \mn@doi [\apj]
  {10.1086/318388}, \href
  {https://ui.adsabs.harvard.edu/abs/2001ApJ...547..792D} {547, 792}

\bibitem[\protect\citeauthoryear{{Fa{\'u}ndez}, {Bronfman}, {Garay}, {Chini},
  {Nyman}  \& {May}}{{Fa{\'u}ndez} et~al.}{2004}]{Faundez2004}
{Fa{\'u}ndez} S.,  {Bronfman} L.,  {Garay} G.,  {Chini} R.,  {Nyman}
  L.~{\r{A}}.,   {May} J.,  2004, \mn@doi [\aap] {10.1051/0004-6361:20035755},
  \href {https://ui.adsabs.harvard.edu/abs/2004A&A...426...97F} {426, 97}

\bibitem[\protect\citeauthoryear{{Fiege} \& {Pudritz}}{{Fiege} \&
  {Pudritz}}{2000}]{Fiege2000}
{Fiege} J.~D.,  {Pudritz} R.~E.,  2000, \mn@doi [\mnras]
  {10.1046/j.1365-8711.2000.03066.x}, \href
  {https://ui.adsabs.harvard.edu/abs/2000MNRAS.311...85F} {311, 85}

\bibitem[\protect\citeauthoryear{{Fuller}, {Williams}  \& {Sridharan}}{{Fuller}
  et~al.}{2005}]{Fuller2005}
{Fuller} G.~A.,  {Williams} S.~J.,   {Sridharan} T.~K.,  2005, \mn@doi [\aap]
  {10.1051/0004-6361:20042110}, \href
  {https://ui.adsabs.harvard.edu/abs/2005A&A...442..949F} {442, 949}

\bibitem[\protect\citeauthoryear{{G{\'o}mez}, {V{\'a}zquez-Semadeni}  \&
  {Zamora-Avil{\'e}s}}{{G{\'o}mez} et~al.}{2018}]{Gomez2018}
{G{\'o}mez} G.~C.,  {V{\'a}zquez-Semadeni} E.,   {Zamora-Avil{\'e}s} M.,  2018,
  \mn@doi [\mnras] {10.1093/mnras/sty2018}, \href
  {https://ui.adsabs.harvard.edu/abs/2018MNRAS.480.2939G} {480, 2939}

\bibitem[\protect\citeauthoryear{{G{\"u}sten}, {Nyman}, {Schilke}, {Menten},
  {Cesarsky}  \& {Booth}}{{G{\"u}sten} et~al.}{2006}]{Gusten2006}
{G{\"u}sten} R.,  {Nyman} L.~{\r{A}}.,  {Schilke} P.,  {Menten} K.,  {Cesarsky}
  C.,   {Booth} R.,  2006, \mn@doi [\aap] {10.1051/0004-6361:20065420}, \href
  {https://ui.adsabs.harvard.edu/abs/2006A&A...454L..13G} {454, L13}

\bibitem[\protect\citeauthoryear{{He} et~al.,}{{He} et~al.}{2015}]{He2015}
{He} Y.-X.,  et~al., 2015, \mn@doi [\mnras] {10.1093/mnras/stv732}, \href
  {https://ui.adsabs.harvard.edu/abs/2015MNRAS.450.1926H} {450, 1926}

\bibitem[\protect\citeauthoryear{{Hennebelle} \& {Andr{\'e}}}{{Hennebelle} \&
  {Andr{\'e}}}{2013}]{Hennebelle2013}
{Hennebelle} P.,  {Andr{\'e}} P.,  2013, \mn@doi [\aap]
  {10.1051/0004-6361/201321761}, \href
  {https://ui.adsabs.harvard.edu/abs/2013A&A...560A..68H} {560, A68}

\bibitem[\protect\citeauthoryear{{Heyminck}, {Kasemann}, {G{\"u}sten}, {de
  Lange}  \& {Graf}}{{Heyminck} et~al.}{2006}]{Heyminck2006}
{Heyminck} S.,  {Kasemann} C.,  {G{\"u}sten} R.,  {de Lange} G.,   {Graf}
  U.~U.,  2006, \mn@doi [\aap] {10.1051/0004-6361:20065413}, \href
  {https://ui.adsabs.harvard.edu/abs/2006A&A...454L..21H} {454, L21}

\bibitem[\protect\citeauthoryear{{Jackson} et~al.,}{{Jackson}
  et~al.}{2019}]{Jackson2019}
{Jackson} J.~M.,  et~al., 2019, \mn@doi [\apj] {10.3847/1538-4357/aaef84},
  \href {https://ui.adsabs.harvard.edu/abs/2019ApJ...870....5J} {870, 5}

\bibitem[\protect\citeauthoryear{{Jim{\'e}nez-Serra}, {Caselli}, {Tan},
  {Hernandez}, {Fontani}, {Butler}  \& {van Loo}}{{Jim{\'e}nez-Serra}
  et~al.}{2010}]{Jimenez2010}
{Jim{\'e}nez-Serra} I.,  {Caselli} P.,  {Tan} J.~C.,  {Hernandez} A.~K.,
  {Fontani} F.,  {Butler} M.~J.,   {van Loo} S.,  2010, \mn@doi [\mnras]
  {10.1111/j.1365-2966.2010.16698.x}, \href
  {https://ui.adsabs.harvard.edu/abs/2010MNRAS.406..187J} {406, 187}

\bibitem[\protect\citeauthoryear{{Klein}, {Philipp}, {Kr{\"a}mer}, {Kasemann},
  {G{\"u}sten}  \& {Menten}}{{Klein} et~al.}{2006}]{Klein2006}
{Klein} B.,  {Philipp} S.~D.,  {Kr{\"a}mer} I.,  {Kasemann} C.,  {G{\"u}sten}
  R.,   {Menten} K.~M.,  2006, \mn@doi [\aap] {10.1051/0004-6361:20065415},
  \href {https://ui.adsabs.harvard.edu/abs/2006A&A...454L..29K} {454, L29}

\bibitem[\protect\citeauthoryear{{Koch} \& {Rosolowsky}}{{Koch} \&
  {Rosolowsky}}{2015}]{Koch2015}
{Koch} E.~W.,  {Rosolowsky} E.~W.,  2015, \mn@doi [\mnras]
  {10.1093/mnras/stv1521}, \href
  {https://ui.adsabs.harvard.edu/abs/2015MNRAS.452.3435K} {452, 3435}

\bibitem[\protect\citeauthoryear{{K{\"o}nyves} et~al.,}{{K{\"o}nyves}
  et~al.}{2015}]{Konyves2015}
{K{\"o}nyves} V.,  et~al., 2015, \mn@doi [\aap] {10.1051/0004-6361/201525861},
  \href {https://ui.adsabs.harvard.edu/abs/2015A&A...584A..91K} {584, A91}

\bibitem[\protect\citeauthoryear{{K{\"o}nyves} et~al.,}{{K{\"o}nyves}
  et~al.}{2020}]{Konyves2020}
{K{\"o}nyves} V.,  et~al., 2020, \mn@doi [\aap] {10.1051/0004-6361/201834753},
  \href {https://ui.adsabs.harvard.edu/abs/2020A&A...635A..34K} {635, A34}

\bibitem[\protect\citeauthoryear{{Krumholz}, {McKee}  \& {Klein}}{{Krumholz}
  et~al.}{2005}]{Krumholz2005}
{Krumholz} M.~R.,  {McKee} C.~F.,   {Klein} R.~I.,  2005, \mn@doi [\nat]
  {10.1038/nature04280}, \href
  {https://ui.adsabs.harvard.edu/abs/2005Natur.438..332K} {438, 332}

\bibitem[\protect\citeauthoryear{{Krumholz}, {Klein}  \& {McKee}}{{Krumholz}
  et~al.}{2007}]{Krumholz2007}
{Krumholz} M.~R.,  {Klein} R.~I.,   {McKee} C.~F.,  2007, \mn@doi [\apj]
  {10.1086/510664}, \href
  {https://ui.adsabs.harvard.edu/abs/2007ApJ...656..959K} {656, 959}

\bibitem[\protect\citeauthoryear{{Lee}, {Myers}  \& {Tafalla}}{{Lee}
  et~al.}{1999}]{Lee1999}
{Lee} C.~W.,  {Myers} P.~C.,   {Tafalla} M.,  1999, in {Nakamoto} T.,  ed.,
  Star Formation 1999. pp 177--178

\bibitem[\protect\citeauthoryear{{Leung} \& {Brown}}{{Leung} \&
  {Brown}}{1977}]{Leung1977}
{Leung} C.~M.,  {Brown} R.~L.,  1977, \mn@doi [\apjl] {10.1086/182446}, \href
  {https://ui.adsabs.harvard.edu/abs/1977ApJ...214L..73L} {214, L73}

\bibitem[\protect\citeauthoryear{{Li}, {Zhang}, {Pillai}, {Stephens}, {Wang}
  \& {Li}}{{Li} et~al.}{2019}]{Li2019}
{Li} S.,  {Zhang} Q.,  {Pillai} T.,  {Stephens} I.~W.,  {Wang} J.,   {Li} F.,
  2019, \mn@doi [\apj] {10.3847/1538-4357/ab464e}, \href
  {https://ui.adsabs.harvard.edu/abs/2019ApJ...886..130L} {886, 130}

\bibitem[\protect\citeauthoryear{{Liu}, {Ho}  \& {Zhang}}{{Liu}
  et~al.}{2010}]{Liu2010}
{Liu} H.~B.,  {Ho} P. T.~P.,   {Zhang} Q.,  2010, \mn@doi [\apj]
  {10.1088/0004-637X/725/2/2190}, \href
  {https://ui.adsabs.harvard.edu/abs/2010ApJ...725.2190L} {725, 2190}

\bibitem[\protect\citeauthoryear{{Liu}, {Wu}, {Wu}, {Qin}  \& {Zhang}}{{Liu}
  et~al.}{2013}]{Liu2013}
{Liu} T.,  {Wu} Y.,  {Wu} J.,  {Qin} S.-L.,   {Zhang} H.,  2013, \mn@doi
  [\mnras] {10.1093/mnras/stt1650}, \href
  {https://ui.adsabs.harvard.edu/abs/2013MNRAS.436.1335L} {436, 1335}

\bibitem[\protect\citeauthoryear{{Liu} et~al.,}{{Liu} et~al.}{2016a}]{Liu2016b}
{Liu} T.,  et~al., 2016a, \mn@doi [\apj] {10.3847/0004-637X/824/1/31}, \href
  {https://ui.adsabs.harvard.edu/abs/2016ApJ...824...31L} {824, 31}

\bibitem[\protect\citeauthoryear{{Liu} et~al.,}{{Liu} et~al.}{2016b}]{Liu2016}
{Liu} T.,  et~al., 2016b, \mn@doi [\apj] {10.3847/0004-637X/829/2/59}, \href
  {https://ui.adsabs.harvard.edu/abs/2016ApJ...829...59L} {829, 59}

\bibitem[\protect\citeauthoryear{{Liu} et~al.,}{{Liu} et~al.}{2020a}]{Liu2020}
{Liu} T.,  et~al., 2020a, \mn@doi [\mnras] {10.1093/mnras/staa1577}, \href
  {https://ui.adsabs.harvard.edu/abs/2020MNRAS.496.2790L} {496, 2790}

\bibitem[\protect\citeauthoryear{{Liu} et~al.,}{{Liu} et~al.}{2020b}]{Liu2020b}
{Liu} T.,  et~al., 2020b, \mn@doi [\mnras] {10.1093/mnras/staa1501}, \href
  {https://ui.adsabs.harvard.edu/abs/2020MNRAS.496.2821L} {496, 2821}

\bibitem[\protect\citeauthoryear{{Louvet} et~al.,}{{Louvet}
  et~al.}{2016}]{Louvet2016}
{Louvet} F.,  et~al., 2016, \mn@doi [\aap] {10.1051/0004-6361/201629077}, \href
  {https://ui.adsabs.harvard.edu/abs/2016A&A...595A.122L} {595, A122}

\bibitem[\protect\citeauthoryear{{Marsh} et~al.,}{{Marsh}
  et~al.}{2016}]{Marsh2016}
{Marsh} K.~A.,  et~al., 2016, \mn@doi [\mnras] {10.1093/mnras/stw301}, \href
  {https://ui.adsabs.harvard.edu/abs/2016MNRAS.459..342M} {459, 342}

\bibitem[\protect\citeauthoryear{{Marsh} et~al.,}{{Marsh}
  et~al.}{2017}]{Marsh2017}
{Marsh} K.~A.,  et~al., 2017, \mn@doi [\mnras] {10.1093/mnras/stx1723}, \href
  {https://ui.adsabs.harvard.edu/abs/2017MNRAS.471.2730M} {471, 2730}

\bibitem[\protect\citeauthoryear{{McKee} \& {Tan}}{{McKee} \&
  {Tan}}{2003}]{McKee2003}
{McKee} C.~F.,  {Tan} J.~C.,  2003, \mn@doi [\apj] {10.1086/346149}, \href
  {https://ui.adsabs.harvard.edu/abs/2003ApJ...585..850M} {585, 850}

\bibitem[\protect\citeauthoryear{{McMullin}, {Waters}, {Schiebel}, {Young}  \&
  {Golap}}{{McMullin} et~al.}{2007}]{McMullin2007}
{McMullin} J.~P.,  {Waters} B.,  {Schiebel} D.,  {Young} W.,   {Golap} K.,
  2007, in {Shaw} R.~A.,  {Hill} F.,   {Bell} D.~J.,  eds,  Astronomical
  Society of the Pacific Conference Series Vol. 376, Astronomical Data Analysis
  Software and Systems XVI. p.~127

\bibitem[\protect\citeauthoryear{{Milam}, {Savage}, {Brewster}, {Ziurys}  \&
  {Wyckoff}}{{Milam} et~al.}{2005}]{Milam2005}
{Milam} S.~N.,  {Savage} C.,  {Brewster} M.~A.,  {Ziurys} L.~M.,   {Wyckoff}
  S.,  2005, \mn@doi [\apj] {10.1086/497123}, \href
  {https://ui.adsabs.harvard.edu/abs/2005ApJ...634.1126M} {634, 1126}

\bibitem[\protect\citeauthoryear{{Molinari} et~al.,}{{Molinari}
  et~al.}{2010}]{Molinari10}
{Molinari} S.,  et~al., 2010, \mn@doi [\aap] {10.1051/0004-6361/201014659},
  \href {https://ui.adsabs.harvard.edu/abs/2010A&A...518L.100M} {518, L100}

\bibitem[\protect\citeauthoryear{{Molinari} et~al.,}{{Molinari}
  et~al.}{2016}]{Molinari16}
{Molinari} S.,  et~al., 2016, \mn@doi [\aap] {10.1051/0004-6361/201526380},
  \href {https://ui.adsabs.harvard.edu/abs/2016A&A...591A.149M} {591, A149}

\bibitem[\protect\citeauthoryear{{Motte}, {Bontemps}  \& {Louvet}}{{Motte}
  et~al.}{2018}]{Motte2018}
{Motte} F.,  {Bontemps} S.,   {Louvet} F.,  2018, \mn@doi [\araa]
  {10.1146/annurev-astro-091916-055235}, \href
  {https://ui.adsabs.harvard.edu/abs/2018ARA&A..56...41M} {56, 41}

\bibitem[\protect\citeauthoryear{{Mueller}, {Shirley}, {Evans}  \&
  {Jacobson}}{{Mueller} et~al.}{2002}]{Mueller2002}
{Mueller} K.~E.,  {Shirley} Y.~L.,  {Evans} Neal~J. I.,   {Jacobson} H.~R.,
  2002, \mn@doi [\apjs] {10.1086/342881}, \href
  {https://ui.adsabs.harvard.edu/abs/2002ApJS..143..469M} {143, 469}

\bibitem[\protect\citeauthoryear{{Myers}, {Mardones}, {Tafalla}, {Williams}  \&
  {Wilner}}{{Myers} et~al.}{1996}]{Myers1996}
{Myers} P.~C.,  {Mardones} D.,  {Tafalla} M.,  {Williams} J.~P.,   {Wilner}
  D.~J.,  1996, \mn@doi [\apjl] {10.1086/310146}, \href
  {https://ui.adsabs.harvard.edu/abs/1996ApJ...465L.133M} {465, L133}

\bibitem[\protect\citeauthoryear{{Nguyen-Lu'o'ng} et~al.,}{{Nguyen-Lu'o'ng}
  et~al.}{2013}]{Nguyen2013}
{Nguyen-Lu'o'ng} Q.,  et~al., 2013, \mn@doi [\apj]
  {10.1088/0004-637X/775/2/88}, \href
  {https://ui.adsabs.harvard.edu/abs/2013ApJ...775...88N} {775, 88}

\bibitem[\protect\citeauthoryear{{Ohlendorf}, {Preibisch}, {Gaczkowski},
  {Ratzka}, {Ngoumou}, {Roccatagliata}  \& {Grellmann}}{{Ohlendorf}
  et~al.}{2013}]{Ohlendorf2013}
{Ohlendorf} H.,  {Preibisch} T.,  {Gaczkowski} B.,  {Ratzka} T.,  {Ngoumou} J.,
   {Roccatagliata} V.,   {Grellmann} R.,  2013, \mn@doi [\aap]
  {10.1051/0004-6361/201220218}, \href
  {https://ui.adsabs.harvard.edu/abs/2013A&A...552A..14O} {552, A14}

\bibitem[\protect\citeauthoryear{{Peretto} et~al.,}{{Peretto}
  et~al.}{2013}]{Peretto2013}
{Peretto} N.,  et~al., 2013, \mn@doi [\aap] {10.1051/0004-6361/201321318},
  \href {https://ui.adsabs.harvard.edu/abs/2013A&A...555A.112P} {555, A112}

\bibitem[\protect\citeauthoryear{{Poglitsch} et~al.,}{{Poglitsch}
  et~al.}{2010}]{Poglitsch2010}
{Poglitsch} A.,  et~al., 2010, \mn@doi [\aap] {10.1051/0004-6361/201014535},
  \href {https://ui.adsabs.harvard.edu/abs/2010A&A...518L...2P} {518, L2}

\bibitem[\protect\citeauthoryear{{Rebolledo} et~al.,}{{Rebolledo}
  et~al.}{2016}]{Rebolledo2016}
{Rebolledo} D.,  et~al., 2016, \mn@doi [\mnras] {10.1093/mnras/stv2776}, \href
  {https://ui.adsabs.harvard.edu/abs/2016MNRAS.456.2406R} {456, 2406}

\bibitem[\protect\citeauthoryear{{Smith}, {Longmore}  \& {Bonnell}}{{Smith}
  et~al.}{2009}]{Smith2009}
{Smith} R.~J.,  {Longmore} S.,   {Bonnell} I.,  2009, \mn@doi [\mnras]
  {10.1111/j.1365-2966.2009.15621.x}, \href
  {https://ui.adsabs.harvard.edu/abs/2009MNRAS.400.1775S} {400, 1775}

\bibitem[\protect\citeauthoryear{{Traficante} et~al.,}{{Traficante}
  et~al.}{2011}]{Traficante2011}
{Traficante} A.,  et~al., 2011, \mn@doi [\mnras]
  {10.1111/j.1365-2966.2011.19244.x}, \href
  {https://ui.adsabs.harvard.edu/abs/2011MNRAS.416.2932T} {416, 2932}

\bibitem[\protect\citeauthoryear{{Vazquez-Semadeni}}{{Vazquez-Semadeni}}{1994}]{Vazquez-Semadeni1994}
{Vazquez-Semadeni} E.,  1994, \mn@doi [\apj] {10.1086/173847}, \href
  {https://ui.adsabs.harvard.edu/abs/1994ApJ...423..681V} {423, 681}

\bibitem[\protect\citeauthoryear{{V{\'a}zquez-Semadeni}, {G{\'o}mez},
  {Jappsen}, {Ballesteros-Paredes}  \& {Klessen}}{{V{\'a}zquez-Semadeni}
  et~al.}{2009}]{Vazquez-Semadeni2009}
{V{\'a}zquez-Semadeni} E.,  {G{\'o}mez} G.~C.,  {Jappsen} A.~K.,
  {Ballesteros-Paredes} J.,   {Klessen} R.~S.,  2009, \mn@doi [\apj]
  {10.1088/0004-637X/707/2/1023}, \href
  {https://ui.adsabs.harvard.edu/abs/2009ApJ...707.1023V} {707, 1023}

\bibitem[\protect\citeauthoryear{{V{\'a}zquez-Semadeni},
  {Gonz{\'a}lez-Samaniego}  \& {Col{\'\i}n}}{{V{\'a}zquez-Semadeni}
  et~al.}{2017}]{Vazquez-Semadeni2017}
{V{\'a}zquez-Semadeni} E.,  {Gonz{\'a}lez-Samaniego} A.,   {Col{\'\i}n} P.,
  2017, \mn@doi [\mnras] {10.1093/mnras/stw3229}, \href
  {https://ui.adsabs.harvard.edu/abs/2017MNRAS.467.1313V} {467, 1313}

\bibitem[\protect\citeauthoryear{{Wu} \& {Evans}}{{Wu} \&
  {Evans}}{2003}]{Wu2003}
{Wu} J.,  {Evans} Neal~J. I.,  2003, \mn@doi [\apjl] {10.1086/377679}, \href
  {https://ui.adsabs.harvard.edu/abs/2003ApJ...592L..79W} {592, L79}

\bibitem[\protect\citeauthoryear{{Wu}, {Evans}, {Shirley}  \& {Knez}}{{Wu}
  et~al.}{2010}]{Wuj2010}
{Wu} J.,  {Evans} Neal~J. I.,  {Shirley} Y.~L.,   {Knez} C.,  2010, \mn@doi
  [\apjs] {10.1088/0067-0049/188/2/313}, \href
  {https://ui.adsabs.harvard.edu/abs/2010ApJS..188..313W} {188, 313}

\bibitem[\protect\citeauthoryear{{Yuan} et~al.,}{{Yuan}
  et~al.}{2017}]{Yuan2017}
{Yuan} J.,  et~al., 2017, \mn@doi [\apjs] {10.3847/1538-4365/aa7204}, \href
  {https://ui.adsabs.harvard.edu/abs/2017ApJS..231...11Y} {231, 11}

\bibitem[\protect\citeauthoryear{{Yue}, {Qin}, {Liu}, {Tang}, {Wu}, {Wang}  \&
  {Zhang}}{{Yue} et~al.}{2021}]{Yue2021}
{Yue} Y.-H.,  {Qin} S.-L.,  {Liu} T.,  {Tang} M.-Y.,  {Wu} Y.,  {Wang} K.,
  {Zhang} C.,  2021, \mn@doi [Research in Astronomy and Astrophysics]
  {10.1088/1674-4527/21/1/14}, \href
  {https://ui.adsabs.harvard.edu/abs/2021RAA....21...14Y} {21, 014}

\bibitem[\protect\citeauthoryear{{Zhang}, {Andr{\'e}}, {Men'shchikov}  \&
  {Wang}}{{Zhang} et~al.}{2020}]{Zhang2020}
{Zhang} G.-Y.,  {Andr{\'e}} P.,  {Men'shchikov} A.,   {Wang} K.,  2020, \mn@doi
  [\aap] {10.1051/0004-6361/202037721}, \href
  {https://ui.adsabs.harvard.edu/abs/2020A&A...642A..76Z} {642, A76}

\bibitem[\protect\citeauthoryear{{Zhou}, {Evans}, {Koempe}  \&
  {Walmsley}}{{Zhou} et~al.}{1993}]{Zhou1993}
{Zhou} S.,  {Evans} Neal~J. I.,  {Koempe} C.,   {Walmsley} C.~M.,  1993,
  \mn@doi [\apj] {10.1086/172271}, \href
  {https://ui.adsabs.harvard.edu/abs/1993ApJ...404..232Z} {404, 232}

\bibitem[\protect\citeauthoryear{{Zinnecker} \& {Yorke}}{{Zinnecker} \&
  {Yorke}}{2007}]{Zinnecker2007}
{Zinnecker} H.,  {Yorke} H.~W.,  2007, \mn@doi [\araa]
  {10.1146/annurev.astro.44.051905.092549}, \href
  {https://ui.adsabs.harvard.edu/abs/2007ARA&A..45..481Z} {45, 481}

\bibitem[\protect\citeauthoryear{{Zucker} \& {Chen}}{{Zucker} \&
  {Chen}}{2018}]{Zucker2018}
{Zucker} C.,  {Chen} H. H.-H.,  2018, \mn@doi [\apj]
  {10.3847/1538-4357/aad3b5}, \href
  {https://ui.adsabs.harvard.edu/abs/2018ApJ...864..152Z} {864, 152}

\makeatother
\end{thebibliography}

\onecolumn

\begin{multicols}{2}
\appendix

\section{}\label{A}
\subsection{Width of the longest filament}

\begin{figure*}
\centering
\includegraphics[scale=0.22]{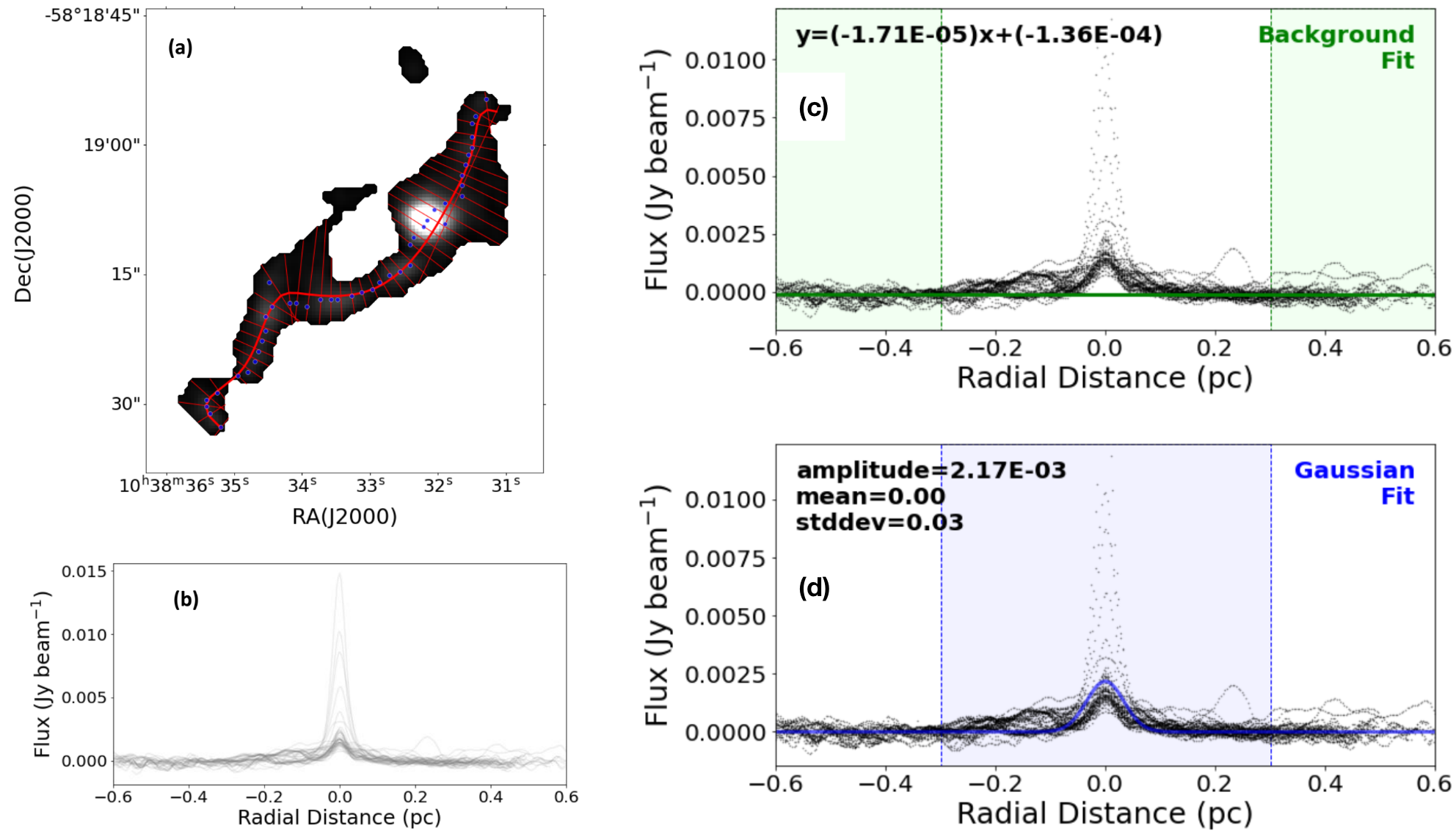}
\caption{(a)The spines of the filaments are marked by the thick red lines in the filament mask. The perpendicular cuts are marked in thin red straight lines. The peak pixel of each intensity cut is indicated by blue dots; (b) The radial distances and density profiles of each cut; (c)The background emission, obtained by fitting over the inner and outer radial distances (bgdegree=1); (d) The result of Gaussian fit, we adopt inner and outer background radii of 0.3 and 0.6 pc. After performing the background subtraction, we fit the Gaussian profile out to a distance of 0.3 pc. The range of radii over which the background and Gaussian profile are fit are highlighted in green and blue regions.}
\label{cut}
\end{figure*}

We employ Radfil algorithm \citep{Zucker2018} to measure the deconvolved FWHM as the width of filament. Based on the mask of filament created by  FILFINDER algorithm \citep{Koch2015}, Radfil can define the spine of the filaments. Along the filament spine, we extract equidistant cuts perpendicular to the smoothed spine, the approximate distance between cuts is 3 pixels. The resulting profiles are shifted such that the center coincides with the peak intensity. When fitting the profiles, we do some background subtraction by adopting the default (bgdegree=1) which specifies the inner and outer radial distances over which to perform the background fit. Then we restrict the fitting range to avoid contamination from peak structures from the other filament.

From Fig.~\ref{cut}(b), we can see a good single-peak profile as a whole.  According to Fig.~\ref{cut}(a), the main deviation comes from the vicinity of G286c1. Based on the star formation history we described in Sec.~\ref{4.3}, the surrounding environment of G286c1 is very complex, which may make it difficult to accurately measure the spine of filament. After performing the background subtraction, we fit the Gaussian profile out to a distance of 0.3 pc (as Fig.~\ref{cut}(c) and (d) shown). The statistical uncertainty on the best-fit amplitude is 0.000034 Jy beam$^{-1}$, and 0.00061 pc for the best-fit mean. Following \citet{Konyves2015}, the deconvolved FWHM = $\sqrt{(FWHM^2 - HPBW^2)}$, where HPBW is the half-power beamwidth, for our observations, the beamwidth of 3mm continuum is $\sim2.20\arcsec$. Finally, we obtain the deconvolved FWHM is $\sim0.075$pc.

\section{}\label{B}

\subsection{Moment maps in ACA 7m array data}

\begin{figure*}
\centering
\includegraphics[scale=0.5]{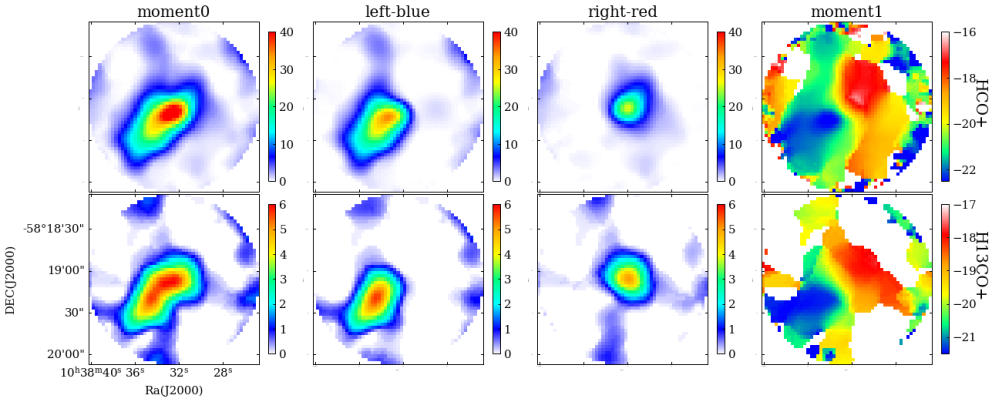}
\caption{Columns from top to bottom show the maps of HCO$^{+}$ (1-0) and H${^{13}}{\rm C}$O$^{+}$ (1-0) with ACA 7m array data. From left to right, the color scales show moment 0 maps of the whole [velocity range is (-60 km s$^{-1}$, 20 km s$^{-1}$)], moment 0 of the left [blueshift, velocity range is (-60 km s$^{-1}$, -19.8 km s$^{-1}$)], moment 0 of the right [redshift, velocity range is (-19.8 km s$^{-1}$, 20 km s$^{-1}$)] and moment 1 of the whole. The color bar on the right indicates the flux scale in Jy beam$^{-1}$ km/s for moment 0 maps and velocity scale in km s$^{-1}$ for moment 1 maps.}
\label{moment7}
\end{figure*}
The left and right columns of Fig.~\ref{moment7} present the moment 0 and moment 1 maps of HCO$^{+}$ (1-0) and H${^{13}}{\rm C}$O$^{+}$ (1-0), respectively. We also integrated the
blue-shifted emission (-60 km s$^{-1}$, -19.8 km s$^{-1}$) and red-shifted emission (-19.8 km s$^{-1}$, 20 km s$^{-1}$) separately and present the integrated intensity maps in the middle two columns of Fig.~\ref{moment7}. From these moment maps, one can see that the G286 clump is composed by two spatially separated sub-clumps with very different velocities. The blueshifted and redshifted gas components are spatially separated. This result indicates that, the double peak profiles of molecular lines in single dish observations are actually caused by gas emission from the two different velocity components (two sub-clumps) rather than gas infall.

\subsection{Molecular line spectra and grid maps in 7m+12m data}

\begin{figure*}
\centering
\includegraphics[scale=0.26]{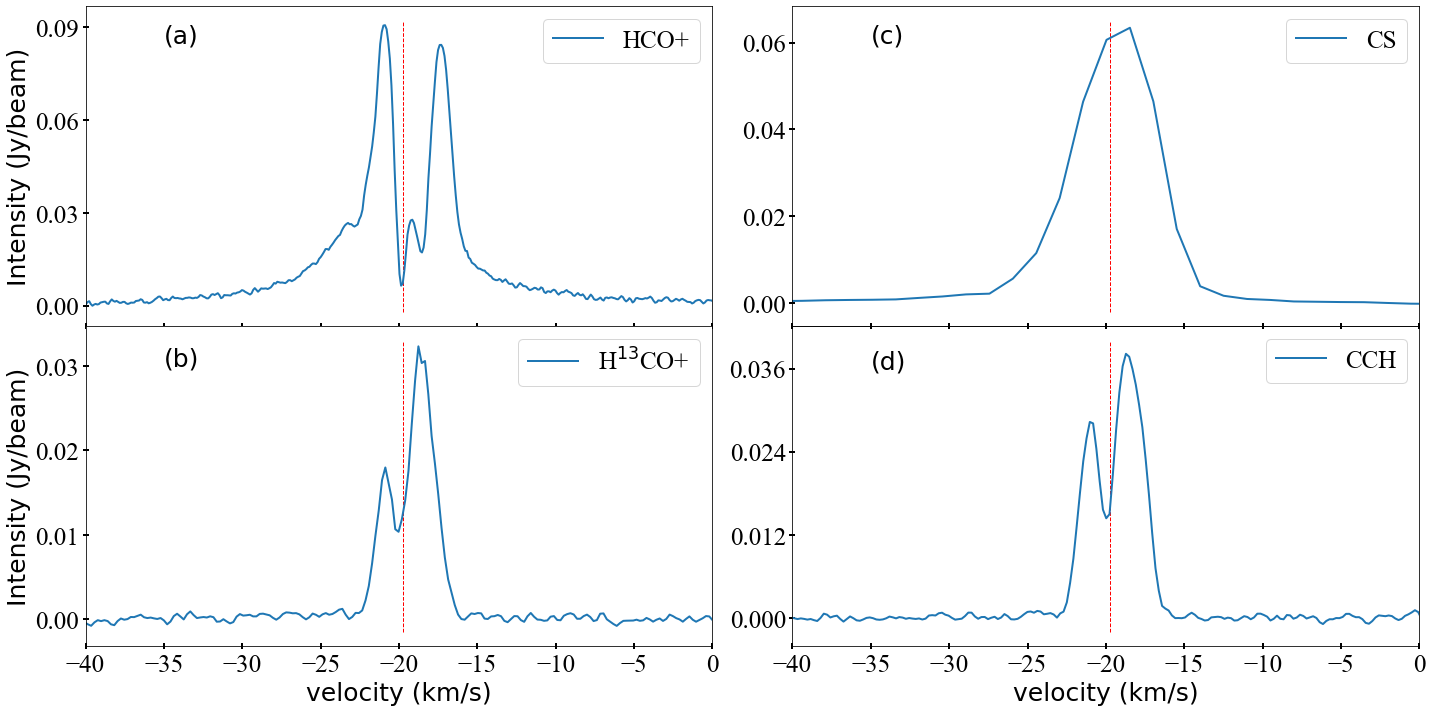}
\caption{(a) HCO$^{+}$ (1-0) spectrum in G286 with 7m+12m data averaged over a region with diameter $36\arcsec$ (the beam size of B10), the vertical dashed red line indicates the systemic velocity at V$_{LSR}$ = -19.8 km s$^{-1}$; (b), (c) and (d) are the averaged spectra of H${^{13}}{\rm C}$O$^{+}$ (1-0), CS (2-1) and CCH (1-0), respectively.}
\label{blue127}
\end{figure*}
\begin{figure*}
\centering
\includegraphics[scale=0.75]{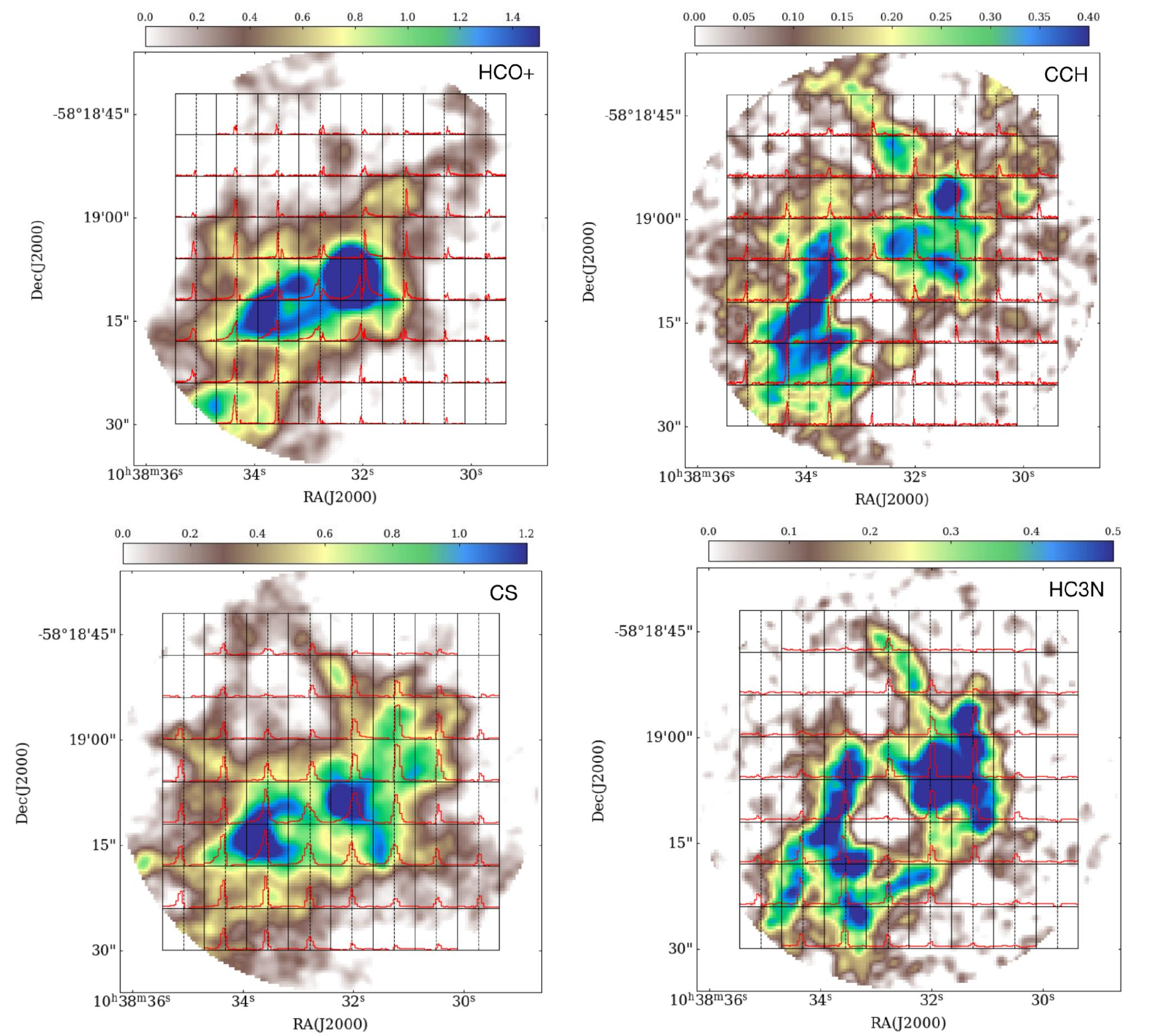}
\caption{The grid maps of HCO$^{+}$ (1-0), CCH (1-0), CS (2-1) and HC$_{3}$N (11-10), background map is the moment 0 maps, the side-length of each square lattice is 15 pixels. The line profile is the averaged spectra in each square lattice of each grid map, the velocity range is (-35 km s$^{-1}$, -5 km s$^{-1}$), black dashed lines in the middle mark the systemic velocity. The color bar on the top indicates the flux scale in Jy beam$^{-1}$ km/s for moment 0 map.}
\label{grid}
\end{figure*}

To confirm the two sub-clumps, we also checked the 12m+7m combined data for HCO$^{+}$ (1-0), H$^{13}$CO$^{+}$ (1-0), CCH (1-0), and CS (2-1). Fig.~\ref{blue127} present their spectra also averaged over an area of 36$\arcsec$ in size. The HCO$^{+}$ and H${^{13}}{\rm C}$O$^{+}$ lines in combined data show very similar line profiles to the spectra in ACA 7m observations.
The line profile of CCH is similar to that of  H${^{13}}{\rm C}$O$^{+}$, showing double peaked profile with the redshifted one stronger than the blueshifted one (or red profile). The average spectra of CS is single-peaked, instead of a double-peak profile. This may be due to the poor spectral resolution ($\sim$1.5 km~s$^{-1}$), which is not high enough to resolve the line profile. Moreover, those features described above are more apparent in the grid maps of Fig.~\ref{grid}.

\section{}\label{C}

\subsection{PV diagrams}

\begin{figure*}
\centering
\includegraphics[scale=0.5]{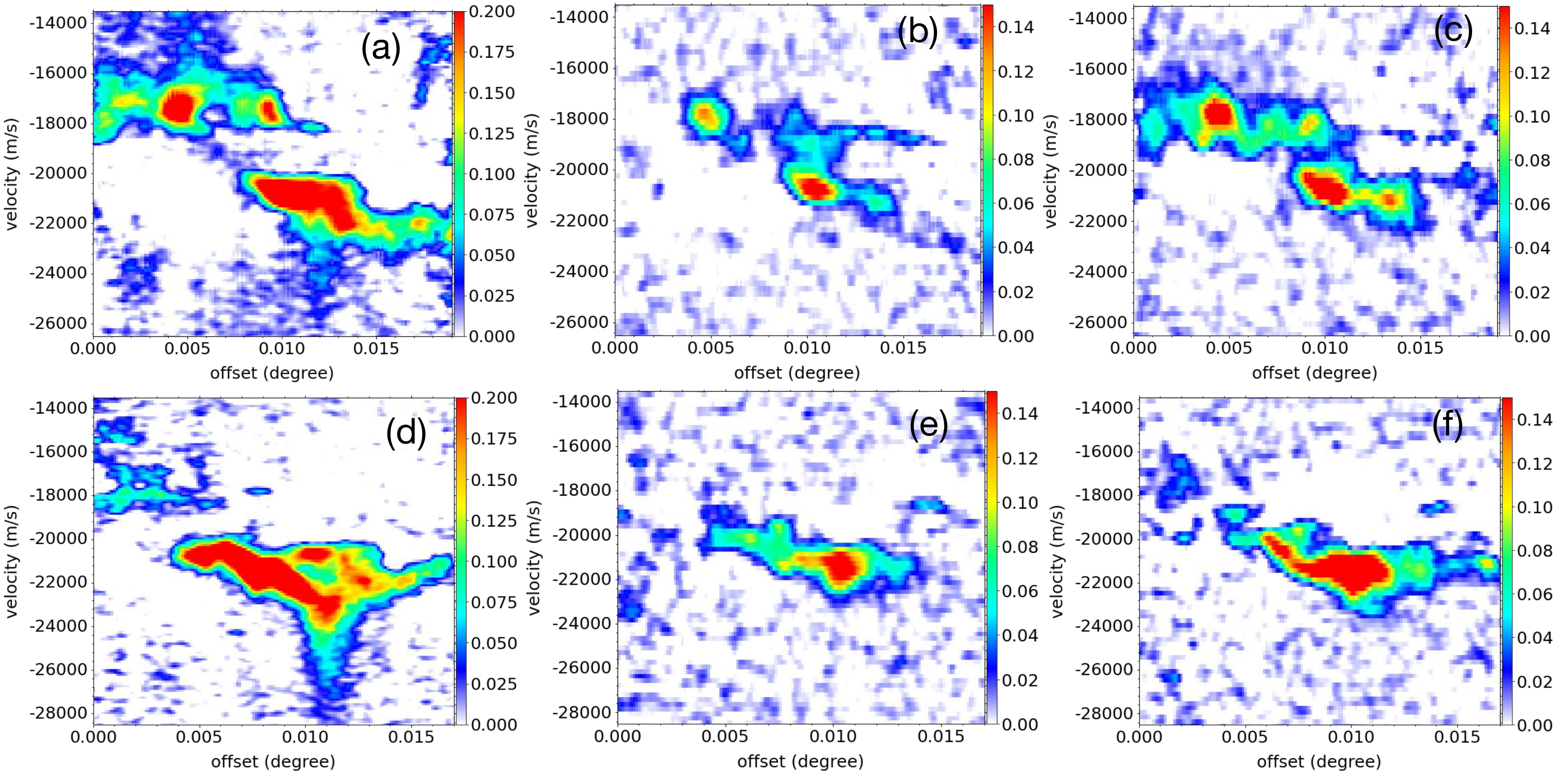}
\caption{PV diagrams of HCO$^+$ J=1-0, H$^{13}$CO$^+$ J=1-0 and CCH J=1-0 (from left to right) along the black and red dashed arrows in Fig.~\ref{big}(a). (a), (b) and (c) are along the black dashed arrow, (d), (e) and (f) for the red dashed arrow.}
\label{pvv}
\end{figure*}

Fig.~\ref{pvv} presents PV diagrams of HCO$^+$ J=1-0, H$^{13}$CO$^+$ J=1-0, and CCH J=1-0 along the two directions marked by long dashed black arrow and red arrow in Fig.~\ref{big}(a), and all of those PV diagrams show distinct velocity gradients similar to that of CS J=2-1 in Fig.~\ref{pv}.

\subsection{Probable interactions between G286 clump, HII region A and HII region B}

\begin{figure*}
\centering
\includegraphics[scale=0.5]{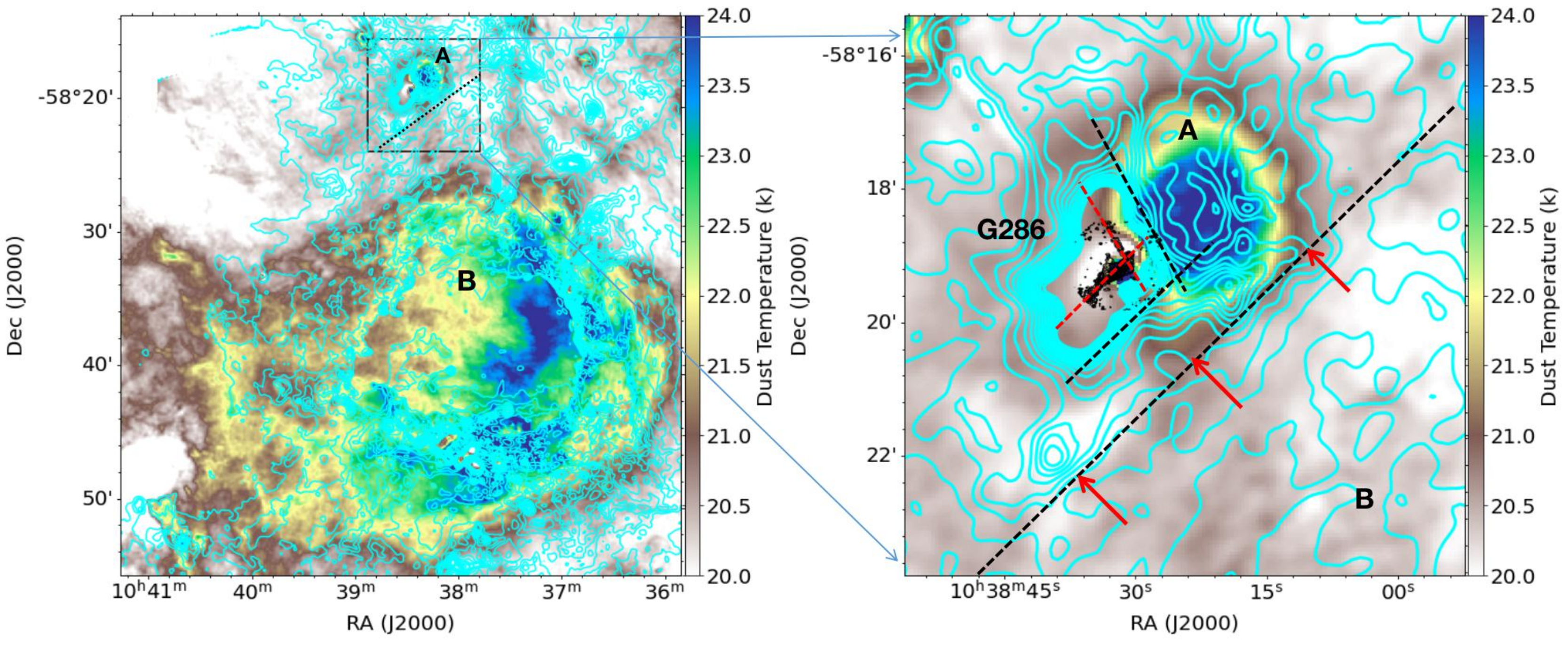}
\caption{Background is the dust temperature map, cyan contour is the column density, all of them derive with the PPMAP procedure \citep{Marsh2017}. Column density contours are from 30 to 200, the step is 10, the unit is 10$^{20}$cm$^{-2}$. Right figure is the blow-up image in the box of left figure, crowded black contours inside G286 clump show the 3 mm continuum emission from 12m+7m data with contour levels 1$\sigma$ $\times$ (1.5, 2, 2.8, 3.6, 4.4 to 12 by 0.4 steps, 12 to 15.2 by 0.8 steps, 18, 24, 30, 36, 48, 60) with $\sigma$=0.21 mJy/beam.}
\label{ppmap}
\end{figure*} 
In Fig.~\ref{ppmap}, the long black dashed line mark the position of density enhancement, red arrow indicates the compression direction. Along the direction of red arrows, the closer to G286 clump, column density contours become the denser. There are no other objects between H{\sc ii} region A and H{\sc ii} region B, and G286 clump. HII region A and HII region B are located in the same region and HII region B is expanding\citep{Cappa2008, Barnes10, Ohlendorf2013, Andersen2017}. Thus the compression is likely from H{\sc ii} region B. From Fig.5 of \citet{Rebolledo2016}, there is a clear velocity gradient due to the expansion of HII region B. Both G286 clump and HII region A are located along the large scale velocity gradient. We can clearly see the velocity gradient from the direction of H{\sc ii} region B in Fig.~\ref{big}(b). Moreover, G286 clump has an obvious morphological deformation in Fig.~\ref{ppmap}, and the southern part of H{\sc ii} region A is obviously dense due to the compression shown in Fig.~\ref{big}(a). In Fig.~\ref{ppmap}, two short black dashed lines mark the dense column density contours. Based on previous analysis, one side is due to the compression of H{\sc ii} region A and the other is likely due to H{\sc ii} region B. Crowded black contours inside G286 clump show the 3 mm continuum emission, two short red dashed lines indicate the two filament in Fig.~\ref{fildc}. Surprisingly, two short red dashed lines are almost parallel to the two short black dashed lines, this strongly implies that the formation of two filaments is due to the compression from two directions described above. 

H{\sc ii} region B is at a certain distance from H{\sc ii} region A and G286 clump. Therefore, we cannot fully determine the compression from H{\sc ii} region B. However, perhaps G286 clump keeps the right distance from H{\sc ii} region B that G286 clump can currently exist as a young massive star formation region, if it's very close to H{\sc ii} region B, the powerful energy injection of H{\sc ii} region B is likely to blow away the matter of G286 clump directly. All of the above speculations need to be further examined.

\end{multicols}

\clearpage

\noindent
Author affiliations:\\
$^{1}$National Astronomical Observatories, Chinese Academy of Sciences, Beijing 100101, Peoples Republic of China \\
$^{2}$University of Chinese Academy of Sciences, Beijing 100049, Peoples Republic of China  \\
$^{3}$Shanghai Astronomical Observatory, Chinese Academy of Sciences, 80 Nandan Road, Shanghai 200030, Peoples Republic of China \\
$^{4}$Key Laboratory for Research in Galaxies and Cosmology, Chinese Academy of Sciences, 80 Nandan Road, Shanghai 200030, Peoples Republic of China \\
$^{5}$Department of Astronomy, Yunnan University, Kunming, 650091, PR China \\
$^{6}$Departamento de Astronom\'ia, Universidad de Concepci\'on, Av. Esteban Iturra s/n, Distrito Universitario, 160-C, Chile \\
$^{7}$Kavli Institute for Astronomy and Astrophysics, Peking University, 5 Yiheyuan Road, Haidian District, Beijing 100871, People's Republic of China\\
$^{8}$Department of Astronomy, Peking University, 100871, Beijing, People's Republic of China\\
$^{9}$Korea Astronomy and Space Science Institute, 776 Daedeokdaero, Yuseong-gu, Daejeon 34055, Republic of Korea\\
$^{10}$University of Science and Technology, Korea (UST), 217 Gajeong-ro, Yuseong-gu, Daejeon 34113, Republic of Korea\\
$^{11}$Physical Research Laboratory, Navrangpura, Ahmedabad—380 009, India \\
$^{12}$Nobeyama Radio Observatory, National Astronomical Observatory of Japan, National Institutes of Natural Sciences, 462-2 Nobeyama, Minamimaki, Minamisaku, Nagano 384-1305, Japan\\
$^{13}$IRAP, Universit\'{e} de Toulouse, CNRS, UPS, CNES, Toulouse, France\\
$^{14}$Institute of Astrophysics, School of Physics and Electronic Science, Chuxiong Normal University, Chuxiong 675000, China.
\noindent

% Don't change these lines
\bsp	% typesetting comment
\label{lastpage}
\end{document}